\documentstyle[epsfig,aps]{revtex}

\def\sla{\slash{\!\!\!} }
\newcommand{\vp}{\mbox{\boldmath $p$}}
\newcommand{\vq}{\mbox{\boldmath $q$}}

\newcommand{\vk}{\mbox{\boldmath $k$}}

\newcommand{\tpsp}{\hspace{1.5em}}
\begin{document}
\title{\bf Spin Polarization and 
       Color Superconductivity in Quark Matter}

\author{E.Nakano$^a$, T.Maruyama$^{b, c, d}$ and T.Tatsumi$^e$}
\address{
         ${^a}${\it Department of Physics, Tokyo Metropolitan University, 
         1-1 Minami-Ohsawa, Hachioji, Tokyo 192-0397, Japan }\\ 
         ${^b}${\it College of Bioresouce Sciences, Nihon University, 
         Fujisawa, 252-8510, Japan} \\
         ${^c}${\it Japan Atomic Energy Research Institute, Tokai, 
                Ibaraki 319-1195, Japan} \\
         ${^d}${\it Institute for Quantum Energy, Nihon University} \\
         ${^e}${\it Department of Physics, Kyoto University, 
                Kyoto 606-8502, Japan} }
\date{\today}
\maketitle

\begin{abstract}
A coexistent phase of spin polarization and color superconductivity 
in high-density QCD is investigated 
using a self-consistent mean-field method at zero temperature.
The axial-vector self-energy stemming from the Fock exchange term 
of the one-gluon-exchange interaction
has a central role to cause spin polarization.
The magnitude of spin polarization is determined 
by the coupled Schwinger-Dyson equations 
with a superconducting gap function. 
As a significant feature,  
the Fermi surface is deformed by the axial-vector self-energy 
and then rotation symmetry is spontaneously broken down. 
The gap function results in being anisotropic in the momentum space 
in accordance 
with the deformation. 
As a result of numerical calculations, 
it is found that spin polarization barely conflicts with color 
superconductivity, 
but almost coexists with it.
\end{abstract}

 
\section{Introduction}
\tpsp
Recently much interest is given for high-density QCD, especially for 
quark Cooper-pair condensation phenomena at high-density quark matter
(called as color superconductivity (CSC)), 
in connection with, e.g., 
physics of heavy ion collisions and neutron stars \cite{CSC1,BL,CSC2}.
Its mechanism is similar to the BCS theory for the 
electron-phonon system \cite{BCS}, 
in which the attractive interaction of electrons is 
provided by phonon exchange and 
causes the Cooper instability near the Fermi surface. 
As for quark matter, 
the quark-quark interaction is mediated by colored gluons, and is 
often approximated by some effective interactions, e.g., 
the one-gluon-exchange (OGE) or the instanton-induced interaction, 
both of which give rise to the attractive quark-quark interaction 
in the color anti-symmetric ${\bf 3^*}$ channel. 
CSC leads to spontaneous symmetry breaking 
of color $SU(3)$ into $SU(2)$ 
as a result of condensation of quark Cooper pairs \cite{BL,CSC2}. 

In this paper we would like to address another phenomenon expected in 
quark matter: spin polarization or ferromagnetism of quark matter.
We examine the possibility of the spin-polarized phase
with CSC in quark matter.
As far as we know, 
interplay between the color superconducting phase 
and other phases characterized 
by the non-vanishing mean fields of the spinor bilinears 
$\langle \bar{\psi} \Gamma \psi \rangle$ has not been explored 
except for the case of chiral symmetry breaking \cite{CSC3}. 
Our main concern here is to investigate the possibility of 
the quark Cooper instability under the axial-vector mean-field, 
$\langle \bar{\psi}\gamma^\mu\gamma_5\psi \rangle$ which is
responsible for 
spin polarization of quark matter.   
It would be worth mentioning in this context that  
ferromagnetism (or spin polarization) and superconductivity are 
fundamental concepts in condensed matter physics, and  
their coexistent phase has been discussed for a long time \cite{MagSup1}.
As a recent progress, a superconducting phase have been discovered 
in ferromagnetic materials and many efforts have been made to understand  
the coexisting mechanism \cite{MagSup2}.

Besides being interesting in its own right, the coexistence problem may be 
related to some physical phenomena.
Recently, a new type of neutron stars, called as ``magnetars'', 
with a super strong magnetic field of $\sim O(10^{15}$G)  
has been discovered \cite{MAG1,MAG2}.
they may raise an interesting question for the origin of the magnetic
field in compact stars,  
since its strength is too large to regard it 
as a successor from progenitor stars, 
unlike canonical neutron stars \cite{MAG3}.
Since hadronic matter spreads over inside neutron stars 
beyond the nuclear density ($\rho_0$$\sim$$0.16$${\rm fm^{-3}}$), 
it should be interesting to consider the microscopic origin of the magnetic 
field in magnetars.
In this context, 
a possibility of ferromagnetism in quark matter 
due to the OGE interaction 
has been suggested by one of the authors (T.T.) 
within a variational framework \cite{Tatsu};  
a competition between the kinetic and 
the Fock exchange energies gives rise to spin polarization, 
similarly to Bloch's idea for itinerant electrons. 
Salient features of spin polarization
in the relativistic system 
are also discussed in Ref. \cite{Tatsu}.
Thus, 
it might be also interesting to examine the possibility of the 
spin-polarized phase
with CSC in quark matter, in connection with magnetars.

We investigate spin polarization in the color superconducting phase 
by a self-consistent framework, 
in which quark Cooper pairs are formed 
under the axial-vector mean-field.
We shall see that this phenomenon is a manifestation of 
spontaneous breaking of both color $SU(3)$ and rotation symmetries.

We adopt here the OGE interaction as an effective quark-quark interaction. 
Since the Fermi momentum is very large at high density,
asymptotic freedom of QCD implies 
that the interaction between quarks is very weak \cite{perry}.  
So it may be reasonable to think that 
the OGE interaction has a dominant contribution for the quark-quark interaction. 
In the framework of relativistic mean-field theories,  
the axial-vector and 
tensor mean-fields, which stem from the Fock exchange terms,  
$\langle \bar{\psi} \gamma_5 \gamma_\mu \psi \rangle$ and 
$\langle \bar{\psi} \sigma_{\mu \nu} \psi \rangle$, 
may have a central role to split the degenerate single-particle energies of 
the two spin states, and then 
leads to spin polarization,
e.g., see \cite{MaruTatsu} for discussion in nuclear matter.
As for quark matter, 
several types of the color singlet mean-fields appear  
after the Fierz transformation in the Fock exchange terms, 
but we retain only the axial-vector mean-field  
as the origin of spin polarization, 
because the OGE interaction by no means holds the tensor mean-field due to 
chiral symmetry in QCD, 
unlike nuclear matter \cite{MaruTatsu}.
Presence of the axial-vector mean-field deforms the quark Fermi seas
according to their spin degrees of freedom, and thereby the gap 
function should be no more isotropic in the momentum space. We assume
here an anisotropic gap
function $\Delta$ on the Fermi surface by a physical consideration and 
solve the coupled Schwinger-Dyson equations self-consistently 
by way of the Nambu formalism to find 
the axial-vector mean-field $U_A$
and the superconducting gap function $\Delta$. 
Thus we discuss the interplay between spin polarization  
and superconductivity in quark matter.

In Section 2 we give a framework to deal with the present subject.
The explicit structure of the anisotropic gap function $\Delta$ in 
the color,
flavor, and Dirac spaces 
is carefully discussed there and in the Appendix B and Appendix C. 
Numerical results about $U_A$ and $\Delta$ are 
given in Section 3, where phase diagram of spin polarization and color  
superconductivity is given in the mass-baryon number density plane. 
Section 4 is devoted to summary and concluding remarks.   
%
\section{Formalism}
%
\tpsp
In this section we present our formalism to treat CSC 
and spin polarization.
We consider quark matter with  flavor $SU(2)$ and color $SU(3)$ symmetries,
and assume that the interaction action is described by 
the OGE interaction as
\begin{eqnarray}
    I_{int}=-g^2\frac{1}{2}\int{\rm d^4}x \int{\rm d^4}y
 \left[\bar{\psi}(x)\gamma^\mu \frac{\lambda_a}{2} \psi(x)\right]
D_{\mu \nu}(x,y)
 \left[\bar{\psi}(y)\gamma^\nu \frac{\lambda_a}{2} \psi(y)\right], 
\end{eqnarray}
where $\psi$ is the quark field, 
$D_{\mu \nu}(x,y)$ is the gauge boson (gluon) propagator, 
and $\lambda_a=1,2,\cdots,8$ are the $SU(3)$ Gell-Mann matrices.
Using the Nambu formalism \cite{BL,Nambu} the effective action  
is given within the mean-field approximation as
\begin{equation}
 I_{MF}=\frac{1}{2} \int \frac{{\rm d}^4 p}{(2 \pi)^4} 
                \left( \begin{array}{l} 
                          \bar{\psi}(p)   \\
                          \bar{\psi}_c(p) \\
                       \end{array} \right)^T
                  G^{-1}(p)
                \left( \begin{array}{l} 
                          \psi(p)   \\
                          \psi_c(p) \\
                       \end{array} \right) \\
\label{mfield}
\end{equation}
with the inverse quark Green function
\begin{equation}
G^{-1}(p)=\left( \begin{array}{cc}
                      \sla{p}-m+\sla{\mu}+V(p) & 
                      \gamma_0 \Delta^\dagger(p) \gamma_0  \\
                      \Delta(p) & 
                      \sla{p}-m-\sla{\mu}+\overline{V}(p) \\
                          \end{array} \right),        
\label{fullg}
\end{equation}
where $\sla{\mu}=\gamma_0 \mu$ with the chemical potential $\mu$.   
$V$ is a self-energy  
and $\Delta$ is the gap function for the quark Cooper pair; both 
terms $V$ and $\Delta$ should be provided by the Fock exchange 
terms 
of the OGE interaction.
We define here $\psi_c(k)$ and $\overline{V}$ as
\begin{eqnarray}
\psi_c(k) &=& C \bar{\psi}^T(-k),
\\
\overline{V} &\equiv& C V^T C^{-1} 
\end{eqnarray}
with the charge conjugation matrix $C$ 
which is explicitly given by $i\gamma_2 \gamma_0$ in Dirac representation.

The Green function $G(p)$ can be written straightforwardly 
from eq.(\ref{fullg})
as
\begin{equation}
 G(p)=\left( \begin{array}{cc}
                            G_{11}(p) & G_{12}(p) \\
                            G_{21}(p) & G_{22}(p) \\
                            \end{array} \right) \label{fullg2} \\
\end{equation}
with
\begin{eqnarray}
  G_{11}(p) &=& \left[ \sla{p}-m+\sla{\mu}+V(p) 
                   -\gamma_0 \Delta(p)^\dagger \gamma_0 
              \left( \sla{p}-m-\sla{\mu}+\overline{V}(p) \right)^{-1}
                    \Delta(p)    \right]^{-1} 
\label{G11i} \\
  G_{21}(p) &=& -\left( \sla{p}-m-\sla{\mu}+\overline{V}(p) \right)^{-1}
              \Delta(p) G_{11}(p). \label{G21i}
\end{eqnarray}

Following Nambu's argument \cite{Nambu}, we impose 
the self-consistency condition to obtain the Hartree-Fock ground state 
such that the self-energy by the residual interaction, $\Sigma_{Res.}$, vanishes: 
 \begin{equation}
  \Sigma_{Res.}=\Sigma_{M.F.}-\Sigma_{Int.}=0, \label{Res}
 \end{equation}
where $\Sigma_{M.F.}$ is defined by  
 \begin{eqnarray}
 \Sigma_{M.F.}(k) &=& G_0^{-1}(k)-G^{-1}(k) =-\left( 
                            \begin{array}{cc}
                               V(k)   & \gamma_0 \Delta^{\dagger}(k) 
\gamma_0 \\
                            \Delta(k) & \overline{V}(k) \\
                            \end{array} \right)\\
\mbox{with} ~~~
G_0(p)&=&\left[\begin{array}{cc}
( \sla{p}-m+\sla{\mu} )^{-1} & 0 \\ 
0 & (\sla{p}-m-\sla{\mu} )^{-1} \\ 
\end{array}\right], 
\end{eqnarray}
and $\Sigma_{Int.}$ is given by the use of the OGE interaction. 
Within the first-order approximation in $g^2$,  
$\Sigma_{Int.}$ renders
\begin{eqnarray}
\Sigma_{Int.}(k)&=&g^2 \int \frac{{\rm d}^4 p}{i(2 \pi)^4}
           D^{a b}(k-p) \hat{\Gamma}_a G(p) \hat{\Gamma}_b \label{self11} \\
 \hat{\Gamma}_a &\equiv& \left( \begin{array}{cc}
                            \gamma^\mu \frac{\lambda_\alpha}{2} & 0 \\
                            0 &  
                    C\left(\gamma^\mu \frac{\lambda_\alpha}{2}\right)^TC^{-1} \\
                            \end{array} \right) 
=                        \left( \begin{array}{cc}
                            \gamma^\mu \frac{\lambda_\alpha}{2} & 0 \\
                            0 & -\gamma^\mu \frac{\lambda^T_\alpha}{2} \\
                            \end{array} \right),
 \end{eqnarray}
which is nothing else but the 
Fock exchange energy by the OGE interaction.
Using eqs. (\ref{Res}) - (\ref{self11}), we obtain the self-consistent equation
for  $V(k)$ 
by the use of the diagonal component of the full Green function (\ref{G11i}): 
 \begin{eqnarray}
-V(k)=(-ig)^2 \int \frac{{\rm d}^4p}{i(2\pi)^4} [-iD^{\mu \nu}(k-p)] 
      \gamma_\mu \frac{\lambda_\alpha}{2} [-iG_{11}(p)] 
      \gamma_\nu \frac{\lambda_\alpha}{2}  
\label{self1}. 
 \end{eqnarray}
The gap equation is also obtained from the off-diagonal component as
\begin{equation}
  -\Delta(k)=(-ig)^2 \int \frac{{\rm d}^4p}{i(2 \pi)^4} [-iD^{\mu \nu}(k-p)]
               \gamma_\mu \frac{-(\lambda_\alpha)^T}{2}
                 [-iG_{21}(p) ] 
               \gamma_\nu \frac{\lambda_\alpha}{2}.  
\label{gap1} 
\end{equation}
In the following sections, 
we present explicit forms of $V(p)$ and $\Delta(p)$  
and then solve their coupled equations (\ref{self1}) and (\ref{gap1}).
\subsection{Fermion propagator under the axial-vector self-energy}
\tpsp

We, hereafter, take the static approximation for the gauge-boson propagator as 
\begin{eqnarray}
D_{\mu \nu}(q) \approx -\frac{g_{\mu \nu}}{{\vq}^2+M^2} \label{cpl}
\end{eqnarray}
where $M$ is an effective gauge boson mass 
originated from the Debye screening 
$M^2 \sim N_f g^2 \mu^2/(2\pi^2)$ \cite{LeBe}.  

Since typical momentum transfer $|{\bf q}|$ at high density is 
of the order of the chemical potential,
we may further introduce the zero-range approximation \cite{Ripka}
for the propagator as
\begin{eqnarray}
D_{\mu \nu}(q) \approx -\frac{g_{\mu \nu}}{Q^2+M^2},
\label{prop-zero}
\end{eqnarray}
with a typical momentum scale $Q$ of $O(\mu)$.
This approximation corresponds to the Stoner model \cite{Yoshi}, 
which is popular in solid-state physics,
and stands on the same concept of the NJL model \cite{NJL} as well.

To proceed, 
we assume, without loss of generality, that total spin expectation value
is oriented to the 
negative $z$-direction in the spin-polarized phase 
which is caused by the finite axial-vector mean-field along the
$z$-axis
\footnote{We shall see that only the space component of the axial-vector 
mean-field is responsible for spin polarization. We,hereafter, take its direction along 
the $z$-axis without loss of generality.} .
As shown in Ref. \cite{MaruTatsu},
rotation symmetry is spontaneously
broken down in this phase while 
axial symmetry
around the $z$-axis is preserved. Then two Fermi seas of the 
different spin states are deformed accordingly.

Applying the Fierz transformation for the Fock exchange energy term (14)
we can see that 
there appear the color-singlet scalar, pseudoscalar, vector and axial-vector
self-energies (Appendix D).   
In general we must take into account these self-energies in $V$,
$V=U_s+i\gamma_5 U_{ps}+\gamma_\mu U_v^\mu+\gamma_\mu\gamma_5U_{av}^\mu$
with the mean-fields $U_\alpha$. 
  Here we introduce an ansatz: the Femri distribution holds the reflection
symmetry with respect to the $p_x - p_y$ plane, and  only the mean-field
parts $U_s$, $U^0_v$ and $U^3_{av}$ are retained in $V$.  
Later we will see that
the self-consistent solution is obtained with the zero-range approximation 
(\ref{prop-zero}) under this ansatz.

In this paper, furthermore, we disregard the scalar mean-field $U_s$ and 
the time component of the vector mean-field $U_v^0$ for simplicity 
since they are irrelevant for the spin degree of freedom;  
$U_v^0$ has only a role to shift
the total energy ot the chemical potential, and may not affect any other 
physical properties.
On the other hand, $U_s$ may  
significantly influence the spin-polarization properties through changing 
the quark effective mass. 
Instead of introducing the scalar mean-field explicitly, however,
we treat the quark mass as a variable parameter,
and discuss its effect in the next section.

According to the above assumptions and considerations 
the self-energy $V$ in eq.(\ref{fullg}) renders 
\begin{equation}
V = \gamma_3 \gamma_5 U_A, ~~~U_A\equiv U_{av}^3 , 
\end{equation}
with the axial-vector mean-field $U_A$.
Then the diagonal component of the Green function $G_{11}(p)$ is written as
\begin{equation}
  G_{11}(p)=\left[ G_A^{-1}-
              \gamma_0 \Delta^\dagger \gamma_0 \tilde{G}_A \Delta \right]^{-1} 
\end{equation}
with
\begin{eqnarray}
   G_A^{-1}(p) &=& \sla{p}-m+\sla{\mu}-\gamma_5 \gamma_3 U_A, \\
  \tilde{G}_A^{-1}(p) &=& \sla{p}-m-\sla{\mu}-\overline{\gamma_5 \gamma_3} U_A,
\end{eqnarray}
where  $\overline{\gamma_5 \gamma_3}=\gamma_5 \gamma_3$ and 
$G_A(p)$ is the Green function with the axial-vector mean-field $U_A$  
which is determined self-consistently by way of eq.~(\ref{self1}).  

Before constructing the gap function $\Delta$, 
we first find the single-particle spectra 
and their eigenspinors in the absence of $\Delta$, which 
is achieved by diagonalization of the operator $G_A^{-1}$. In the usual case of no 
spin polarization this procedure gives nothing but the free energy spectra 
and plane waves. 
Then we choose a gap structure on the basis of a physical 
consideration as in the usual BCS theory.

From the condition that $\det G_A^{-1}(p_0)|_{\mu=0} =0$
one can obtain four single-particle energies 
$\epsilon_\pm$ (positive energies) and $-\epsilon_\pm$ (negative energies), 
which are given as 
\begin{eqnarray}
&& \epsilon_{\pm}({\vp}) = 
  \sqrt{{\vp}^2 + U_A^2 + m^2 \pm
             2 U_A \sqrt{m^2 + p_z^2 }}, 
\label{eig}
\end{eqnarray}
where the sign factor $\pm 1$ being in front of $U_A$ indicates the energy splitting between  
different spin states due to the presence of the axial-vector self-energy, 
which corresponds to the {\it exchange splitting} in the 
non-relativistic electron system \cite{Yoshi}.
In the following, we call the ``spin''-up (-down) states 
for the states $\pm \epsilon_+$ ($\pm \epsilon_-$).
Eq.~(\ref{eig}) also shows that each Fermi sea for 
the ``spin''-up (-down) state should undergo a deformation and lose
rotation symmetry, once $U_A$ is finite. This is a genuine relativistic 
effect \cite{MaruTatsu}; actually the exchange splitting never produces 
deformation of the Fermi sea in
the non-relativistic ferromagnetism, e.g. in the Stoner model \cite{Yoshi}. 

\begin{figure}[ht]
\epsfxsize=13.5cm
\epsfysize=6.8cm
\centerline{\epsffile{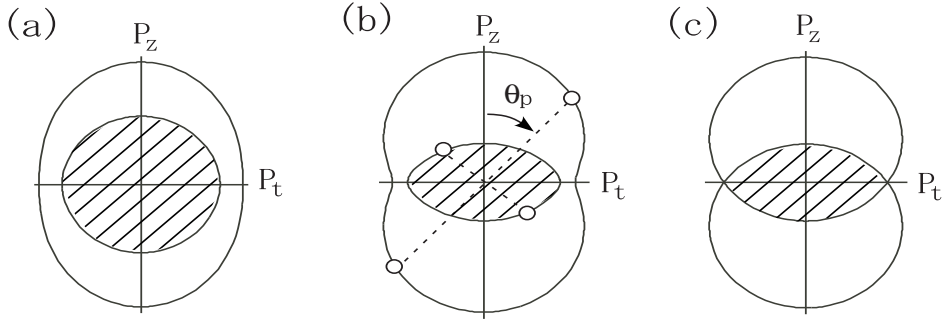}}
\vspace*{-2cm}
\caption{Illustrations of the Fermi surfaces in the $p_t-p_z$ plane 
with $p_t \equiv \sqrt{p_x^2+p_y^2}$. 
{\bf (a)} : for $\mu > U_A+m$ and $U_A \le m$. 
Outer closed curve corresponds to the Fermi surface of the ``spin''-down 
state with single-particle energy $\epsilon_-({\vp})$ 
and inner one surrounding the shaded area  
to the ``spin''-up state with $\epsilon_+({\vp})$.
{\bf (b)} : the same for {\bf (a)} but $U_A \ge m$.
A pair of white circles connected by dashed line represents 
the Cooper pair characterized by $B_n$ (\ref{Bn}). 
Each particle in the Cooper pair has a different color and flavor.
{\bf (c)} : the Fermi surfaces for $m \rightarrow 0$. 
The outer (inner) contour represents 
the Fermi surface for $\epsilon_-$ ($\epsilon_+$). }
\label{FS}
\end{figure}

Here, it would be interesting to see the peculiarities of 
the quark Fermi seas in the presence of the axial-vector self-energy.
In Fig.~{\ref{FS}} we sketch the profile of the Fermi seas projected onto 
the $p_z-p_t$
plane ($p_t = \sqrt{p_x^2 + p_y^2 }$)
for the cases of (a) $U_A < m$, (b) $U_A > m$ and (c) $m=0$.
As is already mentioned 
these seas still hold the axial symmetry 
around the $z$-axis
and the reflection symmetry with respect to the $p_x-p_y$ plane.
The region surrounded by the outer line show the Fermi sea 
of ``spin''-down quarks, 
and the shaded region is that of ``spin''-up quarks. 

We can see in Fig.~{\ref{FS}}a that the Fermi seas 
for the ``spin''-down and ``spin''-up states are deformed in  
the prolate and oblate shapes, respectively, 
where the minimum of the single-particle
energy still resides at the origin ${\vp}=0$.
When $U_A > m$ as shown in Fig.~{\ref{FS}}b, there appear two minima at the points  
$(p_t, p_z) = (0, \pm \sqrt{U_A^2 - m^2})$ for the ``spin''-down quark.
Hence in the massless limit, $m \rightarrow 0$,  
the Fermi sea is described by 
two identical spheres  with radii $\mu$ in the momentum space, which are  
centered at the points $(p_t, p_z) = (0, \pm U_A)$ (see Fig.~{\ref{FS}}c).

In what follows we use subscript `$n$'$(=1,2,3,4)$ for notational convenience 
as $\epsilon_n$ which means 
$\{\epsilon_1, \epsilon_2, \epsilon_3, \epsilon_4 \}=\{\epsilon_-,\epsilon_+,-\epsilon_-,-\epsilon_+\}$.
We define the spinor $\phi_n({\vp})$ that satisfies the equation 
$G_A^{-1}(p_0=\epsilon_n-\mu)\phi_n({\vp})=0$, which corresponds to the 
eigenspinor with the single-particle energy $\epsilon_n$ in the absence 
of the quark Cooper pairing.
The spinor $\phi_n({\vp})$ is explicitly given as 
\begin{equation}
 \phi_n({\vp})={\cal N}_n \left(
  \begin{array}{c}
   (\epsilon_n -(-1)^n \beta_p -U_A) (p_x-{\rm i} p_y) p_z \\
   -((-1)^n \beta_p +m ) p_t^2\\
   \left[ -((-1)^n \beta_p +m)(\epsilon_n-m-U_A)+p_z^2 \right](p_x-{\rm i} p_y)\\  
   p_t^2 p_z
  \end{array} \right), \label{spinor}
\end{equation}
where 
${\cal N}_n$$=\sqrt{[\beta_p-(-1)^n m]\left[ \epsilon_n+U_A+(-1)^n \beta_p \right]/
                           (\epsilon_n \beta_p)}/(2p_t^2 p_z)$
and $\beta_p \equiv \sqrt{p_z^2+m^2}$.
It is to be noted that the spinors $\phi_n$ do not return to 
the eigenspinors of spin operator $\sigma_z$ even when $U_A \rightarrow 0$,
but become mixtures of them, see Appendix A.
Introducing the projection operator  
$\Lambda_n=\phi_n \phi_n^\dagger$ 
with properties  
$\Lambda_{m} \Lambda_n =\Lambda_n \delta_{m n}$ and  
$\sum_n \Lambda_n = {\bf 1}$,
we can recast $G_A(p)$ in the spectral representation into 
\begin{eqnarray}
 G_A &=& \sum_n \frac{\Lambda_n}{p_0-\epsilon_n+\mu} \gamma_0 \\
 G_A^{-1} &=& \sum_n (p_0-\epsilon_n+\mu) \gamma_0 \Lambda_n.
\end{eqnarray}
%
\subsection{Gap structure}
%
In this subsection, 
we give the explicit form of the gap function $\Delta$ 
in the Dirac, color and flavor spaces, and then 
calculate the diagonal component of the full Green function $G_{11}(p)$ 
in eq.~(\ref{G11i}), 
provided that only the axial-vector self-energy is taken 
for $V(p)$ in eq.~(\ref{self1}). In general various types of the 
gap structures are 
possible in 
the Dirac, color and flavor spaces; they depend on the form of interaction 
and the quark mass \cite{BL,QMass}, especially on the strange quark mass \cite{SMass}. 
Here we suppose a simple gap structure from a physical consideration,
disregarding the finite mass effect. 

Using the the spinor $\phi_n({\vp})$ 
we assume that the gap function $\Delta$ 
in eq.~(\ref{gap1}) has a following form in the color and flavor spaces: 
\begin{eqnarray}
\Delta({\vp})=\sum_n \tilde{\Delta}_n({\vp}) B_n({\vp}) \label{aa}
\end{eqnarray}
with the operator $B_n({\vp})$,
\begin{equation} 
B_n({\vp})=\gamma_0 \phi_{-n}({\vp}) \phi_{n}^\dagger({\vp}) \label{Bn}.
\end{equation} 
where the subscript `$-n$'($=-1,-2,-3,-4$) indicates that 
the single-particle energy in the spinor is replaced by that of opposite sign, 
$\epsilon_{-n} \equiv -\epsilon_n$, 
without change of ``spin''.

One can easily see what kind of quark pairs the gap function 
$\Delta$ (\ref{aa}) represents.
Utilizing the property, 
$\phi^T_{-n'}(-{\vp}) C\gamma_0 \phi_{n}({\vp})$$\propto$$\delta_{n' n}$, 
one can find  
for the general spinor 
$\psi({\vp})=\sum_n a_n({\vp}) \phi_{n}({\vp})$ 
with arbitrary coefficients $a_n$, 
\begin{eqnarray}
\bar{\psi}_{c} B_n \psi = 
\psi^T(-p) C \gamma_0 \phi_{-n} \phi^\dagger_{n} \psi(p) \propto
a_{n}(-{\vp})a_{n}({\vp}).
\end{eqnarray} 
This equation clearly shows that two quarks included in the Cooper pairing 
have opposite momenta to each other 
and belong to the same energy eigenstate 
as illustrated in Fig.~\ref{FS}b. 


Now we should note that 
the antisymmetric nature of the fermion self-energy imposes a constraint 
on the gap function \cite{BL,PiRi},
\begin{eqnarray}
C \Delta({\vp}) C^{-1}=\Delta^T({-\vp}).
\end{eqnarray}
Since $B_n$ satisfies the relation 
$C B_n({\vp}) C^{-1} = B_n^T({- \vp})$, \, 
$\tilde{\Delta}_n({\vp})$ must be a symmetric matrix 
in the spaces of internal degrees of freedom. 
Taking into account the property that the most attractive channel of 
the OGE interaction is  
the color antisymmetric ${\bf 3^*}$ one, it must be the flavor singlet 
state.
Thus we can  choose the form of the gap function as 
\begin{equation}   
\left[\tilde{\Delta}_n({\vp})\right]_{\alpha \beta, ~i j}=
\epsilon_{\alpha \beta 3} \epsilon_{i j} \Delta_n({\vp}), \label{2SC}
\end{equation}
where $(\alpha \beta)$ and $(i j)$ are indices in three-color and two-flavor 
spaces, respectively. 
The form of gap function (\ref{2SC}) in the color and flavor spaces is familiar for two-flavor CSC 
\cite{BL,CSC2}.

Using the properties of $\Lambda_n(\vp)$ and $B_n(\vp)$,  
we then obtain an explicit form of $G_{11}(p)$ as
\begin{eqnarray}
  [G_{11}(p)]_{\alpha \beta, i j}
            &=& \left\{ \sum_n \left[ (p_0+\mu-\epsilon_n) -
                 \frac{\Delta_n^\dagger \Delta_n}{p_0+\epsilon_n-\mu} \right] \gamma_0 
                 \Lambda_n \right\}^{-1}_{\alpha \beta, i j} \nonumber \\
            &=& \sum_n \frac{p_0-\mu+\epsilon_n}{
         p_0^2-(\epsilon_n-\mu)^2-\frac{1}{2}{\rm Tr}[\Delta_n^\dagger \Delta_n]
        (1-\delta_{3 \alpha})+i\eta}
       \Lambda_n \gamma_0 \, \delta_{\alpha \beta}\, \delta_{i j} 
 \label{G11} 
\end{eqnarray}
with
\begin{equation}
\Delta_n^\dagger \Delta_n = 
{\rm diag} \left( |\Delta_n|^2, \ |\Delta_n|^2, \ 0 \right)\ \hbox{in the 
color space}, 
\end{equation}
where $\eta$ is a positive infinitesimal.

The quasiparticle energies are obtained by looking for 
the poles of $G_{11}(p)$:
\begin{eqnarray}
E_{n}({\vp})&&=\left\{
 \begin{array}{ll}
 \sqrt{(\epsilon_n({\vp})-\mu)^2+|\Delta_n({\vp})|^2} & \mbox{for color 1, 2} \\
 \sqrt{(\epsilon_n({\vp})-\mu)^2}            & \mbox{for color 3} 
 \end{array} 
        \right.  
\label{qusiE}
\end{eqnarray}
%
The quark number density $\rho_q$ is also given as 
\begin{eqnarray}
\rho_q &\equiv& 
-i \int \frac{{\rm d}^4p}{(2\pi)^4}  
{\rm Tr} \left[(G_{11}(p)- G_{11}(p)|_{\mu=0}) \gamma_0\right]
 \label{BN} \\
& = &
  N_f \sum_{n=1,2} \int \frac{{\rm d}^3 p}{(2\pi)^3}
          \left[ \theta(\mu-\epsilon_{n}) + 2 v_{n}^2({\vp}) 
             - 2 \left(1-v_{-n}^2({\vp})\right)  \right]
 \label{BN2}
\end{eqnarray}  
with 
\begin{equation}
v^2_n({\vp})=\frac{1}{2}\left(1-\frac{\epsilon_n({\vp})-\mu}{E_n({\vp})} \right),  
\label{cof}
\end{equation}
where the first two terms in eq. (\ref{BN2}) show the quark contributions, 
while the last term the anti-quark contribution;
$v_{n}^2({\vp})$ is the occupation probability of the quark pairs   
with momentum ${\vp}$ and represents 
diffuseness of the momentum distribution.

Similarly we can know the self-consistent solutions satisfy 
our ansatz about the mean-fields in $V$. 
From the above solutions we can easily obtaine that 
${\rm Tr}[G_{11}(p) i \gamma_5]=0$,
${\rm Tr}[G_{11}(p)\gamma_i]\propto p_i$,
${\rm Tr}[G_{11}(p)\gamma_5\gamma_0]\propto p_z$ and 
${\rm Tr}[G_{11}(p)\gamma_5\gamma_{1,2}]\propto p_x,p_y$ .
Hence the  pseudoscalar mean-field $U_{ps}$, 
the space-component of vector mean-field $U_v^i$, 
the axial-vector mean-fields $U_{av}^0$ and $U_{av}^{1,2}$ 
are vanished after the integration over angles.

\subsection{Equation for the superconducting gap function}
\tpsp
Using eq.~(\ref{G11}), the off-diagonal component of the full Green 
function $G(p)$, given in eq.~(\ref{G21i}), can be represented in 
the similar way as 
\begin{eqnarray}
  G_{21}(p) =-\sum_n \frac{\gamma_0 B_n \gamma_0}
                          {p_0^2-(\epsilon_n-\mu)^2-|\Delta_n|^2+i\eta}
                          \Delta_n \lambda_2 \tau_2, \label{G21}     
\end{eqnarray}
where $\tau_2$ is the Pauli matrix in the two-flavor space.
Substituting eq.~(\ref{G21}) into the gap equation ~(\ref{gap1}) and 
using the identity  
$\sum_{a=1}^8 (\lambda_a)^T \lambda_2 \lambda_a=-8/3\, \lambda_2$, 
we obtain
\begin{equation}
\sum_{n'} B_{n'}({\vk}) \Delta_{n'}({\vk}) = 
-i\frac{2}{3} g^2 \int \frac{{\rm d}^4 p}{(2\pi)^4} 
  D_{\mu\nu}{(k-p)} \sum_n 
     \left[ \frac{\gamma^\mu \gamma_0 B_n({\vp}) \gamma_0 \gamma^\nu }
             {p_0^2-(\epsilon_n-\mu)^2-|\Delta_n|^2+i\eta} \right]
             \Delta_n({\vp}),                           \label{gap1i}
\end{equation}
where the factor $2/3$ is simply the Fierz coefficient for the color and
flavor degrees of freedom (Appendix~D). 
Furthermore 
multiplying both sides of eq.~(\ref{gap1i}) by $B_{n'}^\dagger({\vk})$ and
taking trace with respect to the Dirac indices,  
the coupled equations for the gap functions $\Delta_n$
are obtained after $p_0$ integration,    
\begin{eqnarray}
 \Delta_{n'}({\vk})=-\frac{2}{3} g^2 \int \frac{{\rm d}^3 p}{(2\pi)^3} 
 D_{\mu\nu}{(k-p)}
 \sum_n T_{n' n}^{\mu \nu}({\vk},{\vp}) 
\frac{\Delta_n({\vp})}{2 E_n({\vp})} \label{GAP1j}  
\end{eqnarray}
where the function $T_{n' n}^{\mu \nu}({\vk}, {\vp})$ is defined as
\begin{eqnarray}
T_{n' n}^{\mu \nu}({\vk},{\vp}) &\equiv& {\rm Tr} 
\left[ B_{n'}^\dagger({\vk}) \gamma^\mu \gamma_0 
        B_n({\vp}) \gamma_0 \gamma^\nu \right] =
(\bar{\phi}_{-n'}({\vk}) \gamma^\mu \phi_{-n}({\vp}))
(\bar{\phi}_{n}({\vp}) \gamma^\nu \phi_{n'}({\vk})) \label{MN},
\end{eqnarray}
a decomposition of $B_n({\vp})$ in terms of gamma matrices and 
its properties are given in Appendix B.

Here we take the zero-range approximation in eq.~(\ref{prop-zero}).
In terms of the polar coordinates ${\vp}=\{p,\theta_p,\phi_p\}$,
we can consider that the gap function $\Delta_n({\vp})$
does not depend on the horizontal angle $\phi_p$ 
due to axial-symmetry around the $p_z$-axis.
Thus we can explicitly perform the integration with respect to 
the angle $\phi_p$ 
in the gap equation (\ref{GAP1j}):
\begin{eqnarray}
 \Delta_{n'}(k,\theta_k)=\frac{2}{3} \tilde{g}^2 
 \int \frac{{\rm d}p\, {\rm d}\theta_p}{(2\pi)^2} p^2 \sin\theta_p
   \sum_n T_{n' n}(k,\theta_k,p,\theta_p) 
   \frac{\Delta_n(p,\theta_p)}{2 E_n(p,\theta_p)} \label{GAP1}  
\end{eqnarray}
with the effective coupling constant $\tilde{g}\equiv g/\sqrt{Q^2+M^2}$.
As seen from the above equation, 
each of the gap functions couples with others  
by the function $T_{n' n}(k,\theta_k,p,\theta_p)$ defined as
\begin{equation}
T_{n' n}(k,\theta_k,p,\theta_p)  \equiv 
\int \frac{{\rm d} \phi_p}{2\pi} g_{\mu \nu}T_{n' n}^{\mu \nu}({\vk},{\vp}) 
= \frac{k_t p_t}{2|\epsilon_{n'}({\vk})| |\epsilon_n({\vp})|}
  \left[ (-1)^{n'+n} \frac{2 m^2+k_z p_z}{\beta_p \beta_k}+1 \right],
\label{Tss1}
\end{equation}
where $p_t$$\equiv$$p\sin\theta_p$ and $p_z$$\equiv$$p\cos\theta_p$ and  
the same for $k_t$ and $k_z$. 
The term proportional to $p_z$ in eq.~(\ref{Tss1}) will disappear 
after the integration over $\theta_p$. 
\subsection{Equation for the axial-vector mean-field $U_A$}
\tpsp
Using eqs.~(\ref{qusiE}) and (\ref{cof}), 
$G_{11}(p)$ is recasted in the form,
\begin{eqnarray}
   &&[G_{11}(p)]_{\alpha \beta, ~i j}
  =\left[\sum_n \left(\frac{1-v_n^2({\vp})}{p_0-E_n+i\eta}+
                              \frac{v_n^2({\vp})}{p_0+E_n-i\eta} \right)
             {\rm e}^{ip_0 \eta} \Lambda_n({\vp}) \gamma_0 \right] 
 \delta_{\alpha \beta} \delta_{i j}. 
\end{eqnarray}
Substituting the above equation into eq.~(\ref{self1}), and 
integrating with respect to $p_0$,
we obtain the self-consistent equation for $U_A$ 
in the zero-range approximation:
\begin{eqnarray}
  U_A
  &=& -\frac{2}{9} \frac{N_f}{2} \tilde{g}^2 \int \frac{{\rm d}^3 p}{(2\pi)^3} 
\sum_n \left[\theta(\mu-\epsilon_n(\vp)) + 2 v_n^2({\vp})\right] S_n({\vp}), 
\label{UA1} 
\end{eqnarray}
where the factor $-2/9$ stems from 
the Fierz coefficient of the color-singlet axial-vector channel of   
the OGE interaction (Appendix D), 
and $S_n({\vp})$ is the expectation value of the spin operator, 
$\sigma_z$$\equiv$$-\gamma_0 \gamma_5 \gamma_3$, 
with respect to the spinor $\phi_n({\vp})$: 
\begin{eqnarray}
  S_n({\vp}) \equiv {\rm Tr} ( \gamma_5 \gamma_3 \Lambda_n({\vp}) \gamma_0 ) = 
  \phi^\dagger_n({\vp}) (-\sigma_z) \phi_n({\vp})=
  \frac{U_A +(-1)^n \beta_p}{\epsilon_n(\vp)}. \label{spin1}
\end{eqnarray}
Thus $U_A$ is related to the expectation value of 
$\sigma_z$ summing over the state with momentum ${\vp}$. 
An effect of the Cooper pairing enters into eq.~(\ref{UA1})
through the function $v_n({\vp})$.
\subsection{Weak coupling approximation}
\tpsp
In this subsection we consider a high-density limit, 
which means the weak coupling limit due to asymptotic freedom of QCD,  
and then disregards the anti-quark pairing and contributions from the  
negative-energy sea (the Dirac sea) in eq.~(\ref{GAP1}). 
Actually  
it costs more energy to form the anti-quark pairing than the quark
pairing for a large chemical potential.  
Taking the approximation, 
we also disregard the contribution from anti-quarks 
to calculate the quark number density in eq.~(\ref{BN2}) and
the axial-vector mean-field eq.~(\ref{UA1}) consistently.
\footnote{This is equivalent to the restriction of the sum over the index  
$n (n=1 - 4)$ to $1,2$, which correspond to the positive-energy states with 
different ``spins'' specified by the subscript $\mp$.} 
In the following calculations we define gap functions of the quark pairing 
by subscript $\pm$ whcih corresponds to the ``spin''-up (-down) of positive-energy states 
as $\Delta_- \equiv \Delta_1$ and $\Delta_+ \equiv \Delta_2$. 
The other symbols with the subscript $\pm$ have the same meaning, 
e.g., $\phi_\mp \equiv \phi_{1,2}$.

In addition, 
we assume that only quarks near the Fermi surface form the Cooper pairs, 
and thereby replace the gap function by an approximated form,
\begin{equation}
\Delta_\pm ({\vp}) \rightarrow \Delta_\pm({\vp}) \theta(\delta-|\epsilon_\pm-\mu|), 
\end{equation}
where 
$\delta$ is a cut-off parameter around the Fermi surface. 
The function $\theta(\delta-|\epsilon_\pm(\vp)-\mu|)$ is also regarded as a form factor to 
regularize the integration in the gap equation \cite{PiRi}. 
The step-function form factor mimics the asymptotic freedom; 
inner particles in Fermi sea costs large kinetic energy to create the pairing and 
takes large momentum transfer which indicates that coupling of this inner-process is small.  
There, however, might be more realistic form factors for finite density QCD, 
which are smoother functions of momentum and $\mu$ than ours, 
we think that they makes little change on qualitative results of the CSC and spin polarization. 
There are models with other form factors or cut-off functions \cite{CSC2,CSC3}.

Looking at the structure of the gap equation (\ref{GAP1}) with (\ref{Tss1}), 
one can find that the gap function is exactly parametrized as (Appendix C)
\begin{eqnarray}
\Delta_\pm({\vp})&=&\frac{p_t}{\epsilon_\pm(\vp)}
\left( \mp\frac{m}{\beta_p} R + F \right) \nonumber\\
&\equiv&\frac{p_t}{\epsilon_\pm(\vp)}\hat\Delta_\pm(\vp) \label{para},  
\end{eqnarray}
where $R$ and $F$ are some constants and represent the 
antisymmetric and symmetric combinations of the gap functions; 
$R=\beta_p/m(\hat\Delta_--\hat\Delta_+)=\beta_p/(p_tm)(\epsilon_-\Delta_- -\epsilon_+\Delta_+)$ 
and $F=\hat\Delta_-+\hat\Delta_+=1/p_t(\epsilon_-\Delta_- +\epsilon_+\Delta_+)$.  
Their magnitudes are determined by the coupled equations; 
\begin{eqnarray}
F&=& \frac{2}{3}\tilde{g}^2
     \int \frac{{\rm d}p\, {\rm d}\theta_p}{(2\pi)^2} p^2 \sin\theta_p
\frac{1}{4}\left[Q_+(\vp) (F - \frac{m}{\beta_p} R) 
                +Q_-(\vp) (F + \frac{m}{\beta_p} R) \right] 
\label{eqF} \\
R&=& \frac{2}{3}\tilde{g}^2
     \int \frac{{\rm d}p\, {\rm d}\theta_p}{(2\pi)^2} p^2 \sin\theta_p
\frac{m}{2\beta_p}\left[-Q_+(\vp) (F - \frac{m}{\beta_p} R) 
                        +Q_-(\vp) (F + \frac{m}{\beta_p} R) \right] 
\label{eqR}, \\ 
&&\mbox{where}~~~ Q_{\pm}(\vp)=
 \frac{p_t^2}{\epsilon_\pm(\vp)^2 E_\pm(\vp)} 
 \theta(\delta-|\epsilon_\pm(\vp) -\mu|).
\nonumber
\end{eqnarray}
We can obviously see that $R\rightarrow 0$ as $m\rightarrow 0$.

\begin{figure}[ht]
\epsfxsize=15cm
\epsfysize=7cm
\centerline{\epsffile{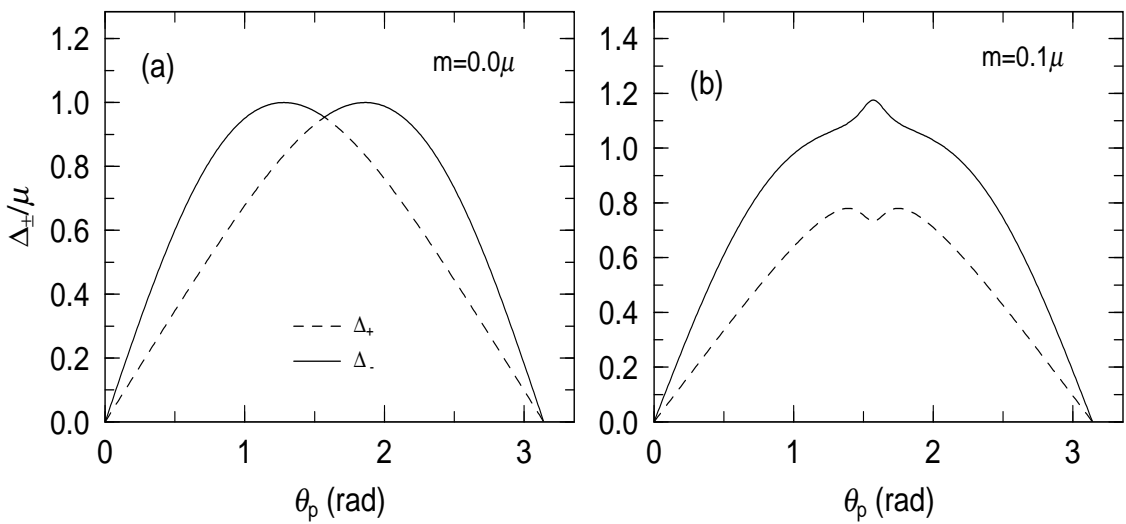}}
\vspace*{0.5cm}
\caption{A schematic view of 
polar-angle dependence of the gap functions at the Fermi surface 
where we set values of the gap parameters as 
$R=0.2\mu$, $F=\mu$, $U_A=0.3\mu$,   
(a) for $m=0$ and (b) for $m=0.1\mu$}
\label{Delta}
\end{figure}

Here we examine the polar-angle dependence of the anisotropic gap function 
at the Fermi surface $\Delta_\pm(p^F,\theta)$.
The Fermi momentum $p^F(\theta)$ of each ``spin'' eigenstate is given as 
\begin{eqnarray}
p_t&=&p^F_\pm(\theta) \sin\theta, ~~p_z=p^F_\pm(\theta) \cos\theta ~~~\mbox{with} 
\nonumber \\
p^F_\pm(\theta)&=&\left[ \mu^2-m^2+U_A^2 \cos(2 \theta) 
\mp U_A \sqrt{4 \mu^2\cos^2\theta+4 m^2\sin^2\theta
-U_A^2\sin^2(2 \theta)} \right]^{1/2}, 
\end{eqnarray}
where the subscript $\pm$ corresponds to the ``spin''-up (-down) state again.
Substituting the above formula 
into the gap function (\ref{para}), we get 
\begin{equation}
\Delta_\pm \left( p^F_\pm,\theta \right)=\frac{p^F_\pm(\theta) \sin \theta}{\mu}
\left[ \mp\frac{m}{\sqrt{m^2+\left(p^F_\pm(\theta) \cos \theta \right)^2}} R + F \right].
\end{equation}
Note that this form exhibits a $P$- wave pairing nature: it is a genuine 
relativistic effect by the Dirac spinors (Appendix B).
We show a schematic view of the above gap functions in Fig.~\ref{Delta}. 

As characteristic features, 
both the gap functions vanish at poles ($\theta=0,\pi$) and take maximal values 
near equator ($\theta=\pi/2$), 
keeping the relation, $\Delta_- \ge \Delta_+$
\footnote{This feature is very similar to $^3P$- pairing in
liquid $^3{\rm He}$ \cite{He3} or nuclear matter \cite{NM3P}.}. 
Suppression of $\Delta_+$ and enhancement of $\Delta_-$ at
$\theta=\pi/2$ for the case of $m\neq 0$ (Fig.~\ref{Delta}b)
are originated from a finite value of $R$, 
while they vanish if quark is taken to be massless (Fig.~\ref{Delta}a). 
The anisotropic gap functions give rise to the different diffuseness 
in the momentum distribution of the two ``spin'' eigenstates, 
and thereby make some effects on spin polarization, unlike in the normal phase.
The anisotropic diffuseness has two effects 
that it obscures the deformation in the momentum distribution 
due to their angle dependence 
and enlarges the difference of the state density between the 
two ``spin'' eigenstates through the relation $\Delta_- \ge \Delta_+$.     
\section{Results and Discussions}
\tpsp
In this section 
we solve the coupled equations (\ref{UA1}), (\ref{eqF}) and (\ref{eqR}) and 
investigate the effects of the superconducting gap on spin polarization.

\begin{figure}[ht]
\epsfxsize=15cm
\epsfysize=6.8cm
\centerline{\epsffile{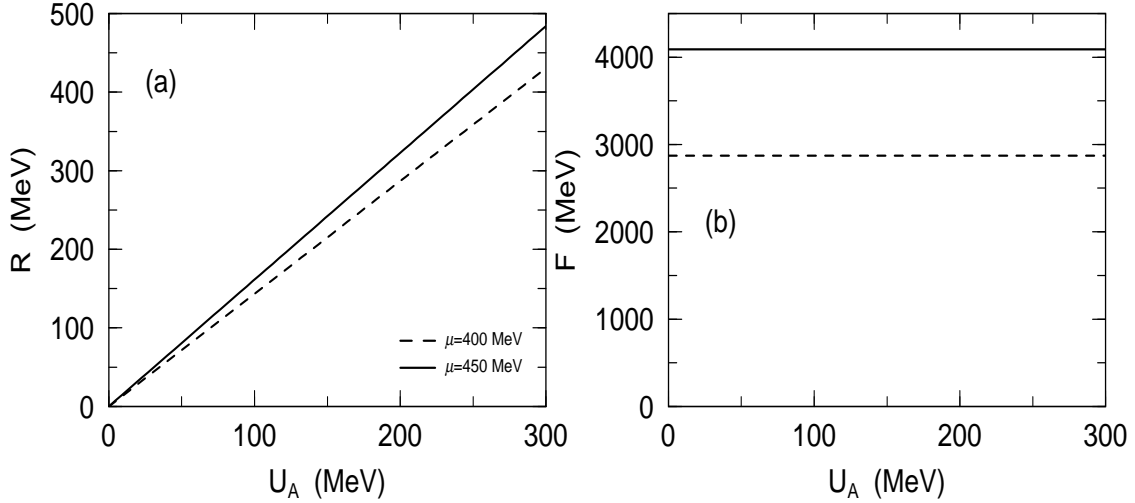}}
\vspace*{0.2cm}
\caption{Parameter dependence of $R$ and $F$ on $U_A$ for 
$\tilde{g}=0.13$ MeV$^{-1}$, $m=20$ MeV and $\delta=0.1\mu$.
{\bf (a)} for $R$ and {\bf (b)} for $F$.
Dashed (solid) lines correspond to $\mu=400 (450)$ MeV. 
The magnitudes of $R$ and $F$ are calculated 
by equations (\ref{eqR}) and (\ref{eqF}) for given $U_A$.}
\label{ParaU}
\end{figure}

Before going to numerical calculations of $U_A$,$R$ and $F$, 
each of which is coupled with others by the self-consistent equations,  
it is instructive 
to see their parameter dependence by treating 
one of them as an input parameter.  
First 
we show $R$ and $F$ as functions of $U_A$ in Fig.~\ref{ParaU}
where $\mu = 400, 450$ MeV and $\delta = 0.1 \mu$.

$R$ starts from zero and almost linearly increases with $U_A$ (Fig.~\ref{ParaU}a), 
which is understood by seeing that 
$R$ is proportional to the difference, ${\hat\Delta}_--{\hat\Delta}_+$,  
due to finite $U_A$, see eq.~(\ref{para}) or Appendix C. 
Thus $R$ is induced by $U_A$ and closely coupled with it. 

As for the behavior of $F$, 
it is barely affected by $U_A$ 
(slight decreasing with $U_A$ in the numerical value) (Fig.~\ref{ParaU}b). 
As seen from the dependence on $\mu$ 
the magnitude of $F$ is almost determined  
by the volume of the phase space in the gap equation,  
that is, by $\mu$ and $\delta$. 
This reflects the fact that $F$ is related to the sum, ${\hat\Delta}_++{\hat\Delta}_-$  
(Appendix C). 
Thus we expect that $F$ increases with density 
when other parameters are fixed.

\begin{figure}[ht]
\epsfxsize=15cm
\epsfysize=6.8cm
\centerline{\epsffile{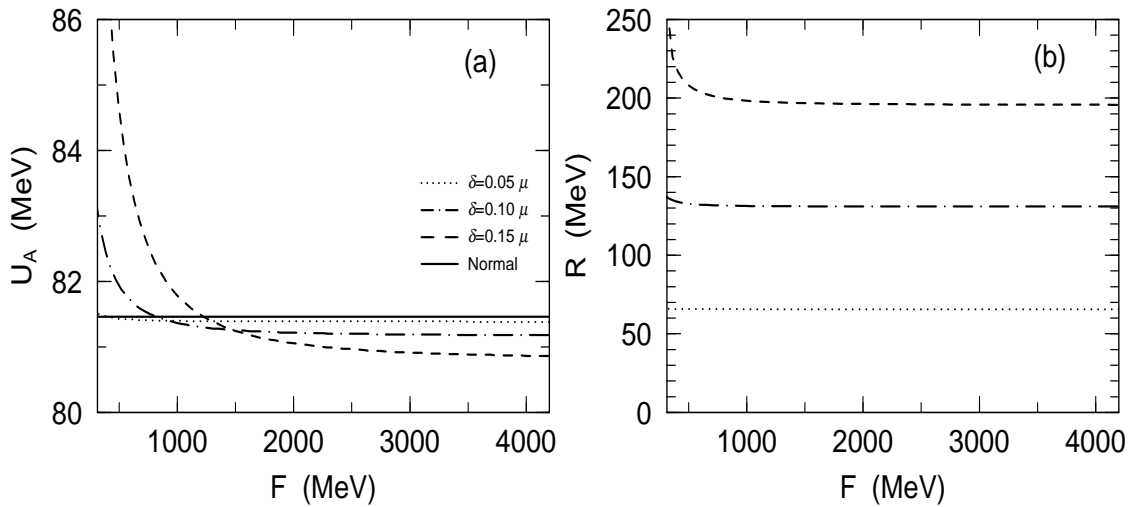}}
\vspace*{0.2cm}
\caption{Parameter dependence of $U_A$ and $R$ on $F$ for 
$\tilde{g}=0.13$ MeV$^{-1}$, $m=20$ MeV and $\mu=450$ MeV.
{\bf (a)} for $U_A$ and {\bf (b)} for $R$.
Solid line, 
dotted lines, dot-dashed lines and dashed lines correspond to 
$\delta=0$ (normal phase), $0.05\mu$, $0.10\mu$ and $0.15\mu$, respectively.
The magnitudes of $U_A$ and $R$ are obtained by their equations for given $F$.}
\label{ParaF}
\end{figure}

From the above results 
we have found that $F$ is not so much influenced by $U_A$.
Next we examine the behavior of $U_A$ and $R$ when $F$ is treated to be 
an input parameter.  
In Fig.~\ref{ParaF}  
we show the parameter dependence of $U_A$ and $R$ on $F$,  
where $\mu=450$ MeV and we use three values of the cut-off parameter 
$\delta =0.05 \mu$, $0.1\mu$ and $0.15\mu$, 
and add the result of $U_A$ in the normal phase ($\delta=0$). 

Comparing the dependence of $U_A$ on $F$ (Fig.~\ref{ParaF}a) 
with that in the normal phase, we see  
a characteristic behavior for different values of $\delta$: 
there are regions 
where $U_A$ is larger than that in the normal phase for relatively small $F$, 
and this region seems to extend with $\delta$.
On the other hand 
results from the self-consistent calculations 
show that $F$ becomes larger with $\delta$ so that 
its value corresponds to a region 
where $U_A$ is comparable with or slightly less  than that 
in the normal phase, 
for any value of the chemical potential.
This situation seems to be qualitatively unchanged, 
once the ratio of the effective coupling constants 
in the axial-vector channel $G_{axial}$ and the diquark channel $G_{diq}$ is kept, 
$G_{axial}:G_{diq}=2/9:2/3$, which comes from the Fierz transformation for color and flavor 
(Appendix D).
However, if the coupling constant in each channel is taken independently,   
our results might be changed qualitatively. 

Seeing the results for $R$ in Fig.~\ref{ParaF}b, 
we find that $R$ increases with $\delta$ due to the growth of the phase space and 
is hardly affected by $F$ 
except the region of small $F$ where $U_A$ varies rapidly as shown 
in Fig.~\ref{ParaF}a: 
it also shows that $R$ is closely related to $U_A$.

These parameter dependences also 
suggest that the regularization scheme for the gap equation, 
i.e.,the sharp momentum cut-off function, the form factor, etc., 
will give rise to a qualitative change to $U_A$. 
In the present cut-off function, $\theta(\delta-|\epsilon_\pm-\mu|)$,
$U_A$ (spin polarization) coexists with CSC, 
except a slight competition, as will be shown later.
\subsection{ Self-consistent solutions}

We demonstrate some self-consistent solutions here.
Since we have little information to determine the values of the parameters 
$\tilde{g}$ and $\delta$ 
(there may be other more reasonable form factors than the present cut-off
function), and 
our purpose is to figure out qualitative properties of spin polarization
in the color superconducting phase,  we mainly set in the following calculations 
them as $\tilde{g}=0.13$ MeV$^{-1}$ and $\delta=0.1\mu$, for example, 
which is not so far from the couplings in NJL-like models \cite{CSC3,Ripka,NJL}.

\begin{figure}[ht]
\epsfxsize=15cm
\epsfysize=6.9cm
\centerline{\epsffile{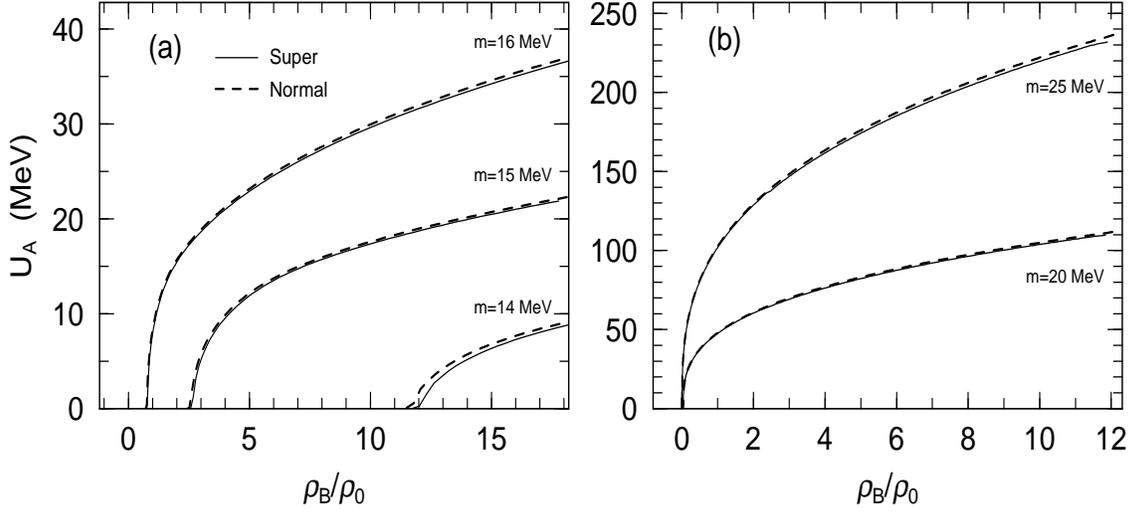}}
\vspace*{0.2cm}
\caption{Axial-vector mean-field as a function of baryon number density
 $\rho_B$ ($\rho_0=0.16$ fm$^{-3}$) 
for $\tilde{g}=0.13$ MeV$^{-1}$ and $\delta=0.1\mu$. 
{\bf (a)} for $m=14 \sim 16$ MeV and {\bf (b)} for $m=20$ and $25$ MeV.
Dashed (Solid) lines are obtained in the normal (color superconducting) phase.}
\label{UA}
\end{figure}

We first examine spin polarization in the absence of CSC.
In Fig.~\ref{UA}   
we show the the axial-vector mean-field $U_A$, 
with $\Delta_\pm $ being set to be zero,
as a function of baryon number density $\rho_B (\equiv \rho_q /3)$ 
relative to the normal nuclear density $\rho_0=0.16$ fm$^{-3}$ 
for $m=14 \sim 25$ MeV (dashed lines). 

It is seen that the axial-vector mean-field (spin polarization) 
appears above a critical density and
becomes larger as baryon number density gets higher.
Moreover, 
the results for different values of the quark mass show that
spin polarization grows more for the larger quark mass.
This is because a large quark mass gives rise to much difference 
in the Fermi seas of two opposite ``spin'' states, 
which leads to growth of the exchange energy in the axial-vector channel. 

\begin{figure}[ht]
\epsfxsize=15cm
\epsfysize=6.9cm
\centerline{\epsffile{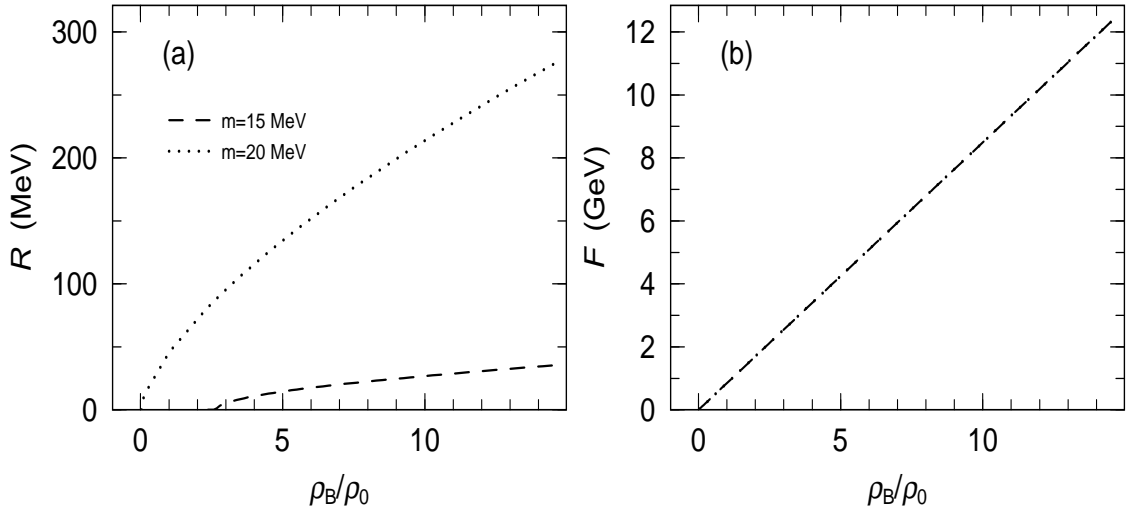}}
\vspace*{0cm}
\caption{$R$ {\bf (a)} or $F$ {\bf (b)} as a function of $\rho_B/\rho_0$ 
for $m=15$ MeV (dashed lines) and $m=20$ MeV (dotted lines).
The other parameters are same in Fig.~\ref{UA}. 
Note that in figure {\bf (b)} the lines almost 
overlap each other for the two quark masses.}
\label{FR}
\end{figure}

\begin{figure}[ht]
\epsfxsize=9cm
\epsfysize=6.8cm
\centerline{\epsffile{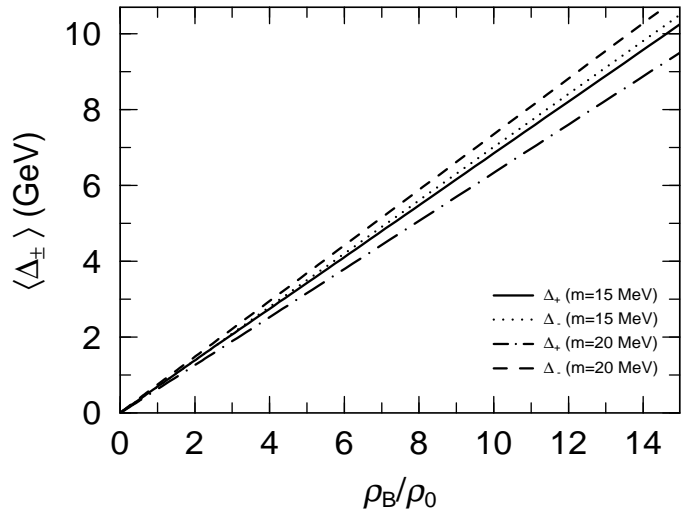}}
\vspace*{0.2cm}
\caption{Mean values of the gap functions with respect to the solid angle 
at the Fermi surface, $\langle \Delta_\pm \rangle$, 
plotted as a function of $\rho_B/\rho_0$ 
for $m=15$ MeV (solid and dotted lines) 
and $20$ MeV (dot-dashed and dashed lines).}
\label{Dpm}
\end{figure}

Next we solve the coupled equations (\ref{UA1}), (\ref{eqF}) and (\ref{eqR}).
Results for $U_A$, $R$ and $F$  are shown in 
Fig.~\ref{UA} (solid lines) and Fig.~\ref{FR}, 
for values of the quark mass  $m=14 \sim 25$ MeV.  
It is found again, by comparing these cases of the quark mass, 
that $U_A$ is very sensitive to the quark mass 
and increases with it as in the absence of CSC (Fig.~\ref{UA}). 
For the behavior of the gap functions, 
$R$ is induced by $U_A$ and 
both of $F$ and $R$ increase with $\rho_B$ 
due to the growth of the Fermi surface (Fig.~\ref{FR}). 
It is also seen that $F$ is not sensitive to the quark mass.
To see the bulk behavior of pairing gap as a function of baryon number
density, 
we also show, in Fig.~\ref{Dpm}, 
their mean values with respect to the polar angle 
on the Fermi surface; 
\begin{equation}
\langle \Delta_\pm \rangle \equiv 
\left( \int_0^\pi {\rm d}\theta \frac{\sin\theta}{2} \Delta_\pm^2 \right)^{1/2}.
\end{equation}
The mean values $\langle \Delta_\pm \rangle$ begin to split with each
other at a density where $U_A$ becomes finite. 
This reflects that $R$ is induced by $U_A$ and then 
has a negative (positive) contribution to $\Delta_+$ ($\Delta_-$). 

Here we would like to comment on the magnitude of $\langle \Delta_\pm \rangle$. 
These should be compared with the usual uniform
gap function, and may look 
very large values of $O$(GeV) in our case.
However these values would be largely reduced by taking a smooth form
factor which models asymptotic freedom of QCD\cite{CSC3};  
it further reduces the integral value in the gap equation, compared with  
our sharp cut-off function.

\begin{figure}[ht]
\epsfxsize=15.7cm
\epsfysize=7.8cm
\centerline{\epsffile{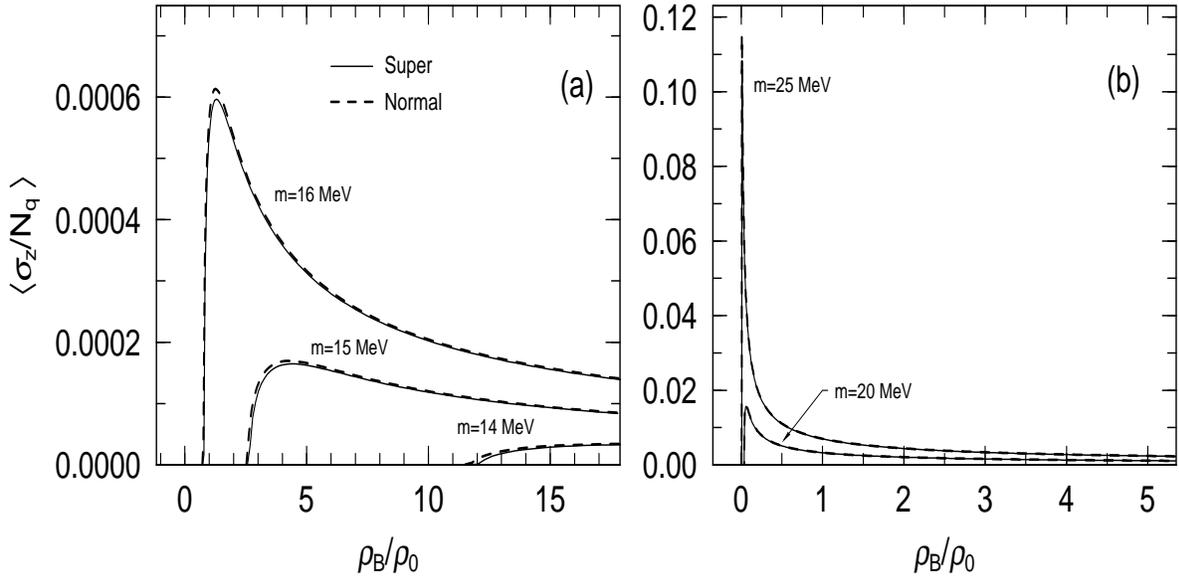}}
\vspace*{0.2cm}
\caption{Spin expectation value per quark as a function of $\rho_B/\rho_0$. 
{\bf (a)} for $m=14 \sim 16$ MeV and {\bf (b)} $m=20$ and $25$ MeV.
Dashed (Solid) lines show results in the normal (color superconducting) phase. }
\label{Spin}
\end{figure}

In Fig.~\ref{Spin} we show the expectation value of the spin operator per quark, 
$\langle \sigma_z/N_q \rangle$, as a function of $\rho_B/\rho_0$ 
with and without the superconducting gap. 
The critical density becomes lower as the quark mass increases, 
and the peak positions of $\langle \sigma_z/N_q \rangle$ 
are located at relatively lower densities in each quark mass.
The magnitude of $\langle \sigma_z/N_q \rangle$ is 
to be compared with $1$ for a free quark, 
because $|\psi_s^\dagger \sigma_z \psi_s/\psi_s^\dagger\psi_s|=1$ 
at the rest frame for the free spinor $\psi_s$.
We arrange the results of three quark masses $m=14\sim 16$ MeV 
by $1$ MeV in Fig.~\ref{Spin}a 
to show a high sensitivity of spin polarization to the quark mass,
which implies that the exchange energy from the attractive axial-vector 
interaction is 
strongly enhanced by the quark mass to produce the large axial-vector 
mean-field. 

The exchange energy is also enhanced by larger chemical potential and 
the resulting axial-vector mean-field increases with it (see Fig.\ref{UA}). 
But the spin expectation value per quark,  
which is relative to the axial-vector mean-field 
per quark ($\propto U_A/N_q$), 
has an upper limit 
since the increase of $N_q$ is far superior to that of $U_A$ 
for larger chemical potential, 
which gives rise to the peak positions in Fig.~\ref{Spin}.   

\begin{figure}[ht]
\epsfxsize=10cm
\epsfysize=6.9cm
\centerline{\epsffile{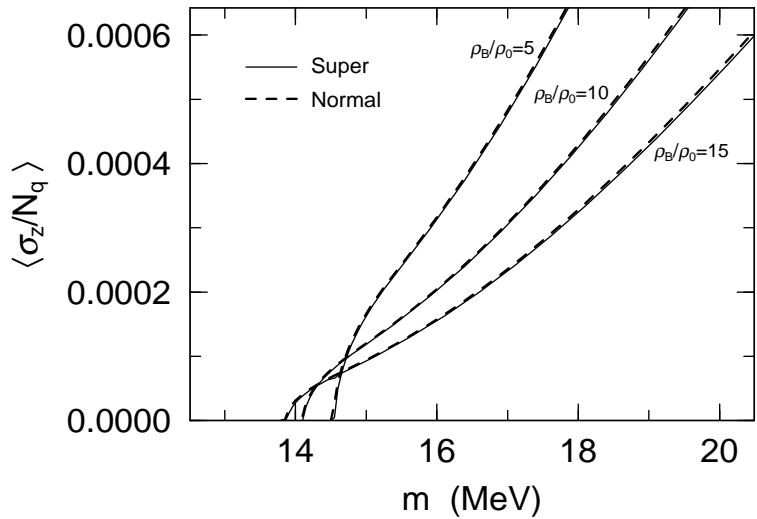}}
\vspace*{0.2cm}
\caption{Spin expectation value per quark as a function of the quark mass $m$ 
for fixed baryon number density $\rho_B/\rho_0=5, 10$ and $15$.}
\label{SpinMass}
\end{figure}

The quark mass is very important in relation to the breaking 
of chiral symmetry in QCD.
Models incorporating chiral dynamics have indicated that
the dynamical mass becomes smaller 
as chiral symmetry is restored at a high density, 
while the current quark mass is small and 
explicitly breaks it \cite{HatsuKuni}.
In our model, on the other hand, 
we treat the quark mass $m$ as a variable parameter 
so that we may simulate a change of the dynamical mass. 
In order to further examine the effect of the quark mass 
on spin polarization, 
we show the mass dependence at densities 
$\rho_B =$ 5$\rho_0$,
$\rho_B =$ 10$\rho_0$  and 
$\rho_B =$ 15$\rho_0$ 
for the cases with and without the superconducting gap 
in Fig.~\ref{SpinMass}. 

Spin polarization increases with the quark mass in all the three densities.
In the figure we exhibit only the results for a narrow region of the mass parameter   
($m=13 \sim 20$ MeV), while  
as for larger masses of $O(100 \mbox{MeV})$ 
(order of the strange quark mass) 
spin polarization monotonically increases without singular oscillations.
Critical values of the quark mass 
at which spin polarization disappears   
become smaller as density increases in both cases.

In relation of $U_A$ to $m$
we can derive an exact result in the massless limit, $m \rightarrow 0$. 
In the normal phase where $\Delta =0$, 
eq.~(\ref{UA1}) becomes
\begin{equation}
U_A
= -\frac{2}{9} \tilde{g}^2 \sum_{n=1,2} \int \frac{{\rm d}^3 p}{(2\pi)^3} 
\, 3 \, \theta(\mu-\epsilon_n({\vp}))
\frac{U_A +(-1)^n |p_z|}{\epsilon_n({\vp})} 
\label{uam0}
\end{equation}
with
\begin{equation}
{\epsilon_\pm({\vp})}  = \sqrt{ \left(|p_z| \pm U_A \right)^2 + p_t^2 } .
\end{equation}
The right-hand side of the above equation can be analytically 
integrated to give
\begin{eqnarray}
U_A
&=& -\frac{2}{9} {\tilde g}^2 \frac{4 \pi}{(2\pi)^3} 
3 \int^{\mu - U_A}_0 d p_z \int^{\sqrt{\mu^2 - (p_z + U_A)^2}}_0
d p_t p_t \frac{U_A + p_z}{\sqrt{ (p_z + U_A)^2 + p_t^2}}
\nonumber \\
&& -\frac{2}{9} {\tilde g}^2 \frac{4 \pi}{(2\pi)^3} 
3 \int^{\mu + U_A}_0 d p_z \int^{\sqrt{\mu^2 - (p_z - U_A)^2}}_0
d p_t p_t \frac{U_A - p_z}{\sqrt{ (p_z - U_A)^2 + p_t^2}}
\nonumber \\
&=& 0
\label{uam01}
\end{eqnarray}
Here we have assumed that $\mu > U_A$.
In the massless limit, 
the Fermi sea is described by two complete spheres 
in the momentum space with radii $\mu$, 
whose centers are located at $(p_t, p_z) = (0, \pm U_A)$ (see Fig.~\ref{FS}c).
The momentum distribution for quarks 
in the ``spin''-down state occupies these two spheres,
while the ``spin''-up  state does their overlap region (shaded).
In the above integration for the ``spin''-down quarks, 
the expectation value of the spin operator given by 
quarks with $0 \le p_z\le U_A$
is canceled with that of quarks with $U_A \le p_z \le 2 U_A$. 
The remaining contribution from the region, 
$2 U_A \le p_z \le \mu + U_A$,  
cancels with that by the ``spin''-up quarks.  
Thus we can see that spin polarization disappears 
as $m \rightarrow 0$ in the absence of CSC.

This analytical result that $U_A \rightarrow 0$ as $m \rightarrow 0$ 
can be also understood as follows. 
The eigenstates of non-interacting massless fermions are classified by
the definite helicity states:
the positive energy state is  
right-handed (left-handed) with positive (negative) helicity, 
while the negative energy state 
those with negative (positive) helicity.
This property is not spoiled by introducing the axial-vector
mean-field, when we extend the meaning of helicity;
the Dirac equations for the "left-" and "right-handed" positive-energy 
fermion fields $\psi_{L,R}$
are now given as 
\begin{eqnarray}
&&(p_0+{\vp}\cdot \mbox{\boldmath $\sigma$}+U_A\sigma_3)\psi_L=0 \\
&&(p_0-{\vp}\cdot \mbox{\boldmath $\sigma$}+U_A\sigma_3)\psi_R=0,
\end{eqnarray}
which give the eigenvalues,
$p_0=\sqrt{p_t^2+(p_z+U_A)^2}(\equiv \epsilon_L({\vp}))$ for $\psi_L$ and
$p_0=\sqrt{p_t^2+(p_z-U_A)^2}(\equiv \epsilon_R({\vp}))$ for $\psi_R$.
$\psi_L$ ($\psi_R$) is the eigenstate of generalized helicity $h=\mp 1$ 
projected onto the shifted momentum ${\vp'}=\{ p_x, p_y, p_z \pm U_A \}$. 
If $\mu \neq 0$ 
they form the spherical Fermi seas, see Fig.~\ref{FS}c.
Here it would be interesting to compare these eigenvalues with the
limit form of $\epsilon_\pm({\vp})$ in eq.~(\ref{eig}),
\begin{equation}
\epsilon_\pm({\vp}) \longrightarrow \sqrt{p_t^2+(|p_z| \pm U_A)^2} ~{\rm as}~m\rightarrow 0.
\end{equation}
Then we can see the relations: 
$\epsilon_\pm({\vp})=\epsilon_L({\vp})\theta(\pm p_z)+\epsilon_R({\vp})\theta(\mp p_z)$, 
which clearly show that the two Fermi seas of the eigenspinors 
(\ref{spinor}) give the same Fermi seas of $\psi_{L,R}$
for a given chemical potential $\mu$. Thus we can take an alternative
view of the Fermi seas in terms of the definite helicity states by rearranging the
eigenspinor (\ref{spinor}) properly in the massless limit.
In each Fermi sea for $\psi_{L,R}$    
the particle number with the definite $h$ becomes the same, and 
thereby the total spin-expectation value becomes vanished. 

In the color superconducting phase, on the other hand, the situation is different  
because the momentum distribution becomes diffused 
due to the creation of the Cooper pairs near the Fermi surface.
For $m \rightarrow 0$ and $\Delta_\pm \neq 0$, 
eq. (\ref{UA1}) becomes 
\begin{equation}
U_A
= -\frac{2}{9} \tilde{g}^2 \sum_{n = 1, 2} \int \frac{{\rm d}^3 p}{(2\pi)^3} 
2 v_n^2({\vp})
\frac{U_A +(-1)^n |p_z|}{\epsilon_n({\vp})} ,
\label{uas}
\end{equation}
where $v_n^2({\vp})$ indicates the diffused part of the momentum distribution, 
defined in eq.~(\ref{cof}). 
Here we should note that the gap functions $\Delta_{\pm}$ are still non zero
even at $m=0$.
The diffused part, however, give no contribution to the spin polarization 
from the viewpoint of the helicity eigenstates. 
The gap function in the massless limit becomes 
\begin{equation}
\Delta_\pm (p,\theta)=
\frac{p_t}{\sqrt{ (|p_z| \pm U_A)^2 + p_t^2 }} F,
\end{equation}
see Fig.~\ref{Delta}a. 
The diffuseness from the above gap function has an equal contribution  
to the two complete Fermi spheres of chirality.
Thus the total spin,  
which is obtained by summing up these momentum distributions,  
should be zero in the massless limit 
even if the CSC is taken into account. 

\begin{figure}[ht]
\epsfxsize=10cm
\epsfysize=6.9cm
\centerline{\epsffile{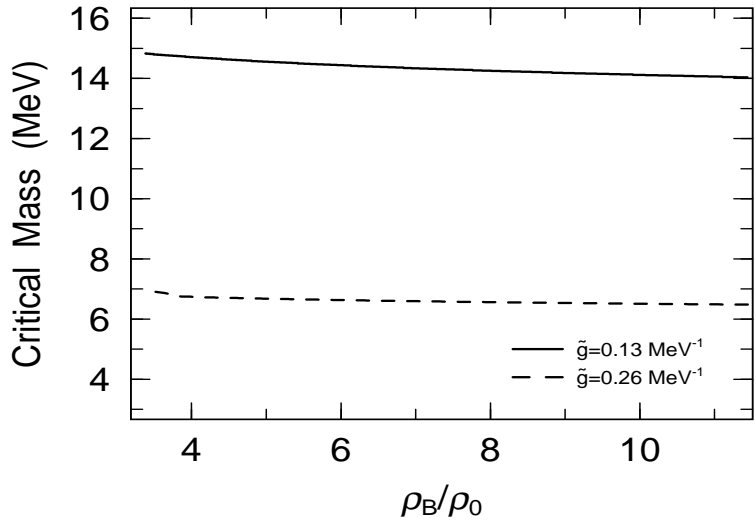}}
\vspace*{0.2cm}
\caption{Phase diagram in $\rho_B/\rho_0 - m$ plane for the effective coupling constant 
$\tilde{g}=0.13$ MeV$^{-1}$ and $0.26$ MeV$^{-1}$. 
At regions above the lines spin polarization arises.
This result is obtained with the gap function, 
thus all the region in the phase diagram shows the color superconducting phase.}
\label{PD}
\end{figure}

To summarize 
we show a phase diagram 
for the quark mass and baryon number density in Fig.~\ref{PD} 
where we add the result of $\tilde{g}=0.26$MeV$^{-1}$ to see the
dependence on the coupling constant. 

The lines indicate the critical mass at a fixed density 
(at regions above the lines spin polarization arises).
We can confirm that the critical mass becomes smaller 
with the increase of the density, 
and spin polarization occurs 
at moderate densities ($\rho_B = 3 \sim 4 \rho_0$) 
if the coupling is strong enough 
even though quark mass is taken to be smaller 
as a simulation for change of dynamical mass (restoration of chiral symmetry).

Here we would like to understand how the gap function affects spin 
polarization 
and brings about a slight reduction of it.
In the spin-polarized phase, 
the momentum distribution is deformed from
the simple spherical shape.
As mentioned in Ref. \cite{MaruTatsu},
the deformation is induced by finite $U_A$ and 
feeds back to $U_A$ in a self-consistent manner. 
In the color superconducting phase, 
diffuseness caused by the Cooper pairing 
in the momentum distribution depends on the polar angle and then
has an influence on the deformation. 
As can be expected from the polar-angle dependence of the gap function, 
diffuseness tends to obscure the deformation.

From the consideration of the spin expectation values 
by spinors (\ref{spinor})
near the Fermi surfaces;
\begin{equation}
\phi_\pm^\dagger (-\sigma_z) \phi_\pm=
\frac{U_A \pm \sqrt{p_z^2+m^2}}{\epsilon_\pm }\approx 
\frac{U_A \pm \sqrt{p_z^2+m^2}}{\mu} \label{NearFermi},
\end{equation}
where $\phi_\mp \equiv \phi_{1, 2}$ for two ``spins''.
The difference of the spin expectation value between two ``spin'' states $\phi_\pm$ 
is largely affected by high-$p_z$ regions or regions 
near both poles ($\theta=0,\pi$).
Thus the large deformation along the $z$-axis 
seems to enhance spin polarization.

\begin{figure}[ht]
\epsfxsize=9cm
\epsfysize=6.7cm
\centerline{\epsffile{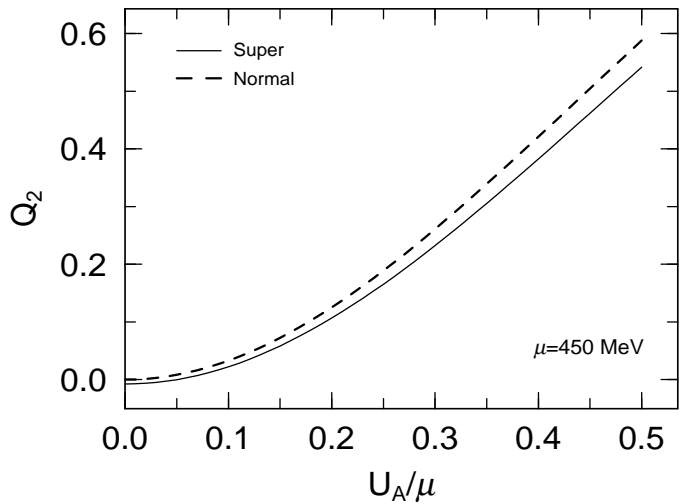}}
\vspace*{0.2cm}
\caption{Quadrupole deformation of the momentum distribution (\ref{quadru}) 
as a function of $U_A$. 
Parameters are fixed as 
$\tilde{g}=0.13$ MeV$^{-1}$, $m=20$ MeV and $\mu=450$ MeV. 
Dashed (solid) line is given in the normal (color superconducting) phase. }
\label{Q2}
\end{figure}

In order to specify to what extent the Fermi sea is deformed, 
we calculate the quadrupole deformation of the momentum distribution defined by 
\begin{equation}
Q_2 \equiv 3\langle p_z^2 \rangle/\langle {\vp}^2 \rangle-1 \label{quadru}.
\end{equation}
In Fig.~\ref{Q2}, 
we show  $Q_2$ as a function of $U_A$ at $\mu=450$ MeV 
in the normal phase 
and in the color superconducting phase 
in which the gap functions are given by their equations for fixed $U_A$. 

From this result of $Q_2$ deformation, 
we can see that 
the diffused part near the Fermi surface obscures the deformation 
then gives an opposite effect against $Q_2$, 
and thus reduces spin polarization. 

\begin{figure}[ht]
\epsfxsize=9cm
\epsfysize=6.7cm
\centerline{\epsffile{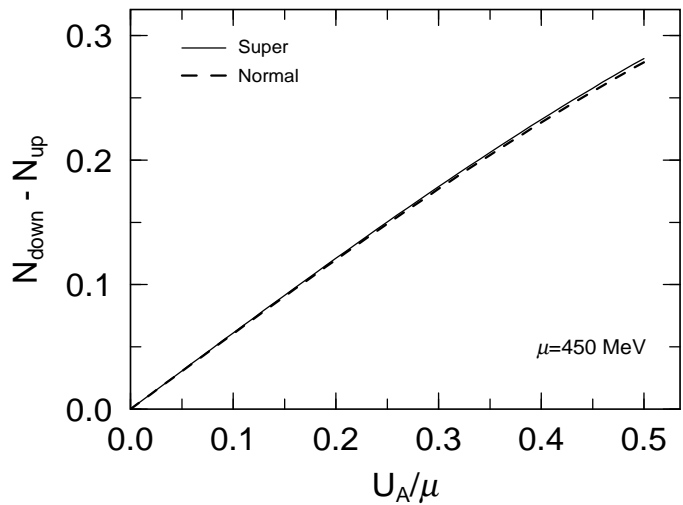}}
\vspace*{0.2cm}
\caption{Difference of the state densities in the two ``spin'' states plotted  
as a function of $U_A$.
Legends are the same as in Fig.~\ref{Q2}.}
\label{Frac}
\end{figure}

Nevertheless the gap function has another effect on spin polarization.
It is to be noted that the qualitative relation, 
$\Delta_- \ge \Delta_+$, is always retained as seen from Fig.~\ref{Delta} 
and then has a effect to enlarge the difference 
of the state density between the two ``spin'' states.
This effect is expected to enhance spin polarization 
since the difference of the spin expectation value by each spinor (\ref{NearFermi}) 
near the equator ($\theta=\pi/2$), so that $p_z\approx0$, 
seems to depend only on the difference of the state density. 
To see it in both the normal and color superconducting phases, 
we define that $N_{up}$ ($N_{down}$) is the state density of the 
``spin''-up (-down) state 
and show their difference by 
$dN \equiv N_{down}-N_{up}$, only in the first two colors, 
as a function of $U_A$ in Fig.~\ref{Frac} at $\mu=450$ MeV. 
The result indicates that the gap functions slightly enhance $dN$ 
than normal phase. 

From the above discussions 
spin polarization is significantly influenced by 
both the deformation and the state density in each ``spin'' state.
As a result of the self-consistent calculation in the color superconducting phase, 
the reduction effect on the deformation is slightly superior to 
the enhancement effect from the difference of the state densities, 
and the pairing effect finally  
reduces spin polarization than in the normal phase.
It, however, should be noted that this qualitative conclusion about 
whether CSC enhances spin polarization 
than normal phase or not is very delicate and may be changed 
depending on the regularization scheme, as already mentioned.
Moreover 
other types of pairing which are not considered here, 
e.g. pairing of the ``spin'' -up and -down states, 
may gives rise to  qualitatively different results, 
while it is very difficult to 
see which type of pairing is energetically favored. 


Finally we would like to comment on the coupling of the spin polarized 
quark matter with the external magnetic field;  
quark fields couple with the magnetic field 
through its anomalous magnetic moment.  
The magnetic interaction is described by the Gordon identity 
for the gauge coupling term: 
$g_L e^*/2m (\bar{\psi} \sigma_{\mu\nu} \psi) F^{\mu\nu}$ 
where $g_L$ is a form factor and $e^*$ an effective charge.  
\footnote{Here we needn't consider the orbital angular momentum 
for uniform matter.
But if a superconductor is of the `second' type 
in which London's penetration length is larger than the coherence length, 
a vortex lattice may be formed in response to the external field 
and then total magnetization is to undergo a qualitative change 
due to circulation of supercurrent \cite{Abri}.}
In quark matter   
a magnetic moment is given as the expectation value
$\langle \sigma_{i j} \rangle$ with respect to the ground state.
In our model 
only $\langle \sigma_{12} \rangle$ is nonzero, and 
the magnetic moment per quark is given as 

\begin{equation}
M_z \equiv \langle \sigma_{1 2}/ N_q \rangle  =
\frac{1}{\rho_q}
\sum_{n=1, 2} \int \frac{d^3p}{(2\pi)^3} [ 2 v_n^2({\vp})+\theta(\mu-\epsilon_n)]
\bar{\phi}_n({\vp}) \sigma_{12} \phi_n({\vp}). 
\label{MZ}
\end{equation}
Note that
the expectation value of $\sigma_{12}$ by the spinor does not depend on $U_A$;  
$\bar{\phi}_\pm ({\vp}) \sigma_{12} \phi_\pm ({\vp})=\mp m/\beta_p$, 
so that $M_z$ reflects only the asymmetry in the momentum distribution 
due to the axial-vector mean-field. 

\begin{figure}[ht]
\epsfxsize=15.5cm
\epsfysize=8cm
\centerline{\epsffile{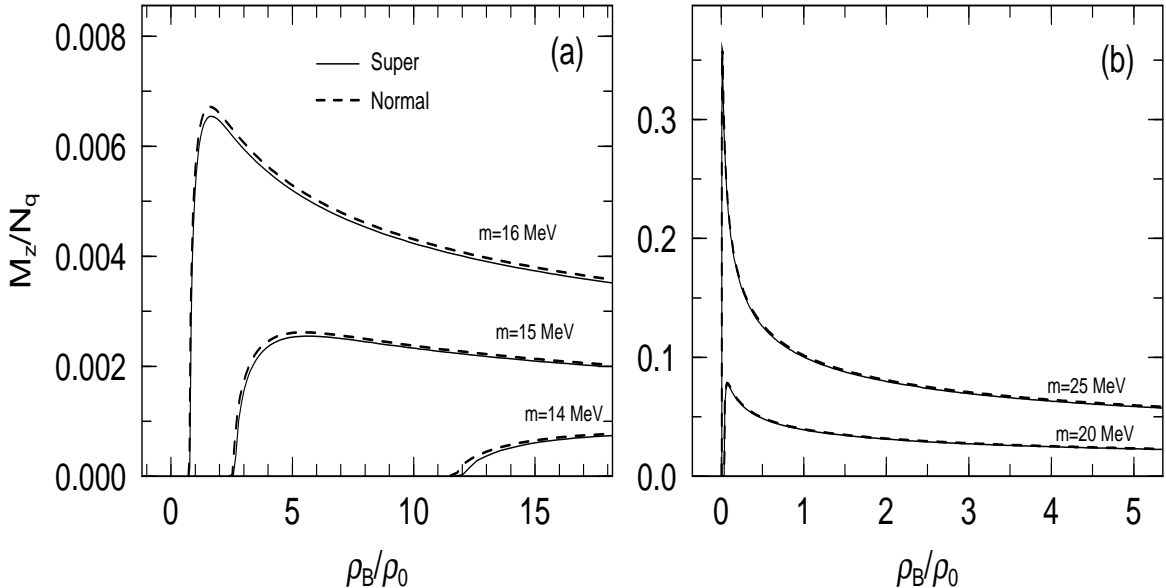}}
\vspace*{0.2cm}
\caption{Induced magnetic moment per quark (\ref{MZ}) as a function of $\rho_B/\rho_0$.
Parameters and legends are the same as in Fig.~\ref{UA}.}
\label{Mag}
\end{figure}

In Fig.~\ref{Mag}, $M_z$ is given 
as a function of baryon number density. 
This indicates that resulting ground state also 
holds ferromagnetism (spontaneous magnetization).
\section{Summary and Concluding remarks}
\tpsp
In this paper 
we have examined spin polarization in quark matter
in the color superconducting phase.
We have introduced the axial-vector self-energy 
and the quark pair field (the gap function), 
whose forms are derived from the one-gluon-exchange interaction 
by way of the Fierz transformation 
under the zero-range approximation. 
Within the relativistic Hartree-Fock framework 
we have evaluated their magnitudes 
in a self-consistent manner by way of the coupled Schwinger-Dyson equations.

As a result of numerical calculations  
spontaneous spin polarization occurs
at a high density for a finite quark mass 
in the absence of CSC, 
while it never appears for massless quarks as an analytical result.
In the spin-polarized phase the single-particle energies corresponding
to 
spin degrees of freedom, which are degenerate in the non-interacting
system,  
are split by the exchange energy in the axial-vector channel. 
Each Fermi sea of the single-particle energy deforms 
in a different way, 
which causes an asymmetry in the two Fermi seas and then  
induces the axial-vector mean-field in a self-consistent manner.   
In the superconducting phase, however, 
spin polarization is slightly reduced by the pairing effect; 
it is caused by 
competition between reduction of the deformation and 
enhancement of the difference in 
the phase spaces of opposite ``spin'' states 
due to the anisotropic diffuseness in the momentum distribution. 

In connection of the deformation with superconductivity 
it has recently been reported \cite{Muther} that in 
the superconducting asymmetric nuclear matter 
the Fermi sea may undergo a deformation even in the spin-saturated system 
due to the difference of the Fermi surface between 
neutrons and protons; 
the momentum distributions of neutrons and protons  
may deform respectively
to enlarge the overlapped region in the phase space, 
which effectively contributes to the $np$- pairing. They have shown the possibility
of the deformation in a variational way; the Fermi sea of 
the majority of nucleons deforms in a prolate shape, while the minority
in an oblate shape. Thus the deformation property of the Fermi seas 
looks very similar 
to our case. Nevertheless, note that our deformation is produced by the relativistic
effect. Anyway it would be interesting to look further into the common feature. 
  
It is to be noted that 
if the effective coupling constant is strong  
enough to lower the critical quark mass,  
spin polarization (magnetization) has potential to appear 
at rather moderate densities such as in the core of neutron stars, 
even though CSC weakly works against it.

From the above observations  
it is suggested that 
spin polarization does not compete with 
CSC but can coexist with it,  
unlike in ordinary superconductors of the electron system with the 
$s$-wave and spin-singlet pairing.
This reflects the fact that internal degrees of freedom of the quark
field, e.g.the  color, flavor and Dirac indices, 
have rich structures 
to satisfy the antisymmetric constraint on the quark-pair field. 

The possibility of the coexistent phase might also 
give a clue as for the origin of the superstrong 
magnetic field observed in magnetars. We roughly estimate the expected
magnetic field when magnetars are assumed to be quark stars. The maximum
dipole magnetic field 
at the star surface reads
\begin{equation}
B_{\rm max}=\frac{8\pi}{3}\mu_qn_q(\langle M_z\rangle/N_q),
\end{equation}
with $\mu_q$ and $n_q$ being the quark magnetic moment and the quark number 
density, respectively;e.g., for $\langle M_z\rangle/N_q\sim (10^{-3})$ and 
$n_q\sim O(1{\rm fm}^{-3})$, we find $B_{\rm max}\sim O(10^{15}{\rm
G})$, which is comparable to that observed in magnetars \cite{MAG1,MAG2}.

In the present paper   
we have not taken into account chiral symmetry, 
which is one of the basic concepts in QCD. 
If chiral symmetry is restored at finite baryon number density,
the quark mass becomes drastically smaller 
as density increases. 
In order to simulate it 
we have examined the quark mass dependence on spin polarization.
In the future work 
we would like to consider an effect of the dynamical mass 
on the axial-vector self-energy.  

{\bf{Acknowledgements.}}

The present
research of T.T. and T.M. is partially supported by the REIMEI Research
Resources of Japan Atomic Energy Research Institute, and by the
Japanese Grant-in-Aid for Scientific Research Fund of the Ministry
of Education, Culture, Sports, Science and Technology (11640272,
13640282).

%
%
\appendix
\section{Structure of spinor under the axial-vector mean-field $U_A$}
In this Appendix 
we rewrite the spinor (\ref{spinor}) 
in terms of the free quark one and the remainder characterized by $U_A$.   
We employ the free spinor $u_s({\vp})$
in which the two-component Pauli spinors 
are given as eigenvectors of the spin matrix  $\sigma_z$:
\begin{eqnarray} 
&& u_\pm({\vp})=  
  \left(
  \begin{array}{c}
   \sqrt{\epsilon_0 +m} \xi_\pm\\
   \frac{\sqrt{\epsilon_0 -m}}{|p|} {\vp} \cdot \mbox{\boldmath $\sigma$} \xi_\pm  
  \end{array} \right) 
~~~\mbox{with}~~~
 \xi_+ =  
  \left(
  \begin{array}{c}
   1\\
   0  
  \end{array} \right)  ~~\mbox{and}~~
 \xi_- =  
  \left(
  \begin{array}{c}
   0\\
   1  
  \end{array} \right),
\end{eqnarray}
where $\epsilon_0=\sqrt{{\vp}^2+m^2}$ 
and $\{ \mbox{\boldmath $\sigma$} \}$  are the Pauli spin matrices.
  
The spinor $\phi_+({\vp})\equiv \phi_2({\vp})$ for the ``spin''-up state is decomposed as follows :
\begin{eqnarray}
&&\frac{ \beta_p({\rm \Delta}\epsilon -2U_A)
   +m(\delta\epsilon +2\beta_p) }{p_t^2 p_z {\cal N}_+} \phi_+({\vp})= \nonumber \\
&& ~~~2\beta_p\sqrt{\epsilon_0+m}
   \left( \frac{\epsilon_+ -\beta_p-U_A}{p_x+ip_y}u_+({\vp})
      -\frac{\beta_p+m}{p_z}u_-({\vp}) \right) 
  +\frac{\mbox{Rem$_1$}(U_A)}{p_z(p_x+ip_y)} ,
\end{eqnarray}
where $\delta\epsilon \equiv \epsilon_- -\epsilon_+$, 
      ${\rm \Delta}\epsilon \equiv \epsilon_- +\epsilon_+$ and 
\begin{equation}
 \mbox{Rem$_1$}(U_A) =
  \left(
  \begin{array}{c}
   p_z(\epsilon_+ -\beta_p-U_A)
   \left[ \beta_p({\rm \Delta}\epsilon -2U_A -2\epsilon_0)
   +m\delta\epsilon \right] \\
  -(p_x+ip_y)(\beta_p+m) 
   \left[ \beta_p({\rm \Delta}\epsilon -2U_A -2\epsilon_0)
   +m\delta\epsilon \right] \\
   \delta\epsilon
   [(\epsilon_+ -\beta_p-U_A)p_z^2 +2\beta_p^2(\beta_p+m)]
   -2\beta_p(\beta_p+m)
   \left[ (\epsilon_- -U_A)(\epsilon_+-U_A) 
   -\epsilon_0^2 \right] \\
   p_z(p_x+ip_y)(\beta_p+m) \delta\epsilon
  \end{array} \right). \nonumber
\end{equation}
Note that the term $\mbox{Rem$_1$}(U_A)$ vanishes in the limit, 
$U_A \rightarrow 0$. 
Thus 
one can fined that $\phi_+({\vp})$ is a mixture of the free spinors  
even when $U_A=0$.

A decomposition of 
the spinor $\phi_-({\vp})\equiv \phi_1({\vp})$ for the ``spin''-down state 
can also be done in the similar way:
\begin{eqnarray}
&&\frac{ \beta_p({\rm \Delta}\epsilon -2U_A)
   +m(\delta\epsilon +2\beta_p) }{p_t^2 p_z {\cal N}_-} \phi_-({\vp})= \nonumber \\
&& ~~~2\beta_p\sqrt{\epsilon_0+m}
   \left( \frac{\epsilon_- +\beta_p-U_A}{p_x+ip_y}u_+({\vp})
         +\frac{\beta_p-m}{p_z}u_-({\vp}) \right) 
  +\frac{\mbox{Rem$_2$}(U_A)}{p_z(p_x+ip_y)},
\end{eqnarray}
where 
\begin{equation}
 \mbox{Rem$_2$}(U_A) =
  \left(
  \begin{array}{c}
   p_z(\epsilon_- +\beta_p-U_A)
   \left[ \beta_p({\rm \Delta}\epsilon -2U_A -2\epsilon_0)
   +m\delta\epsilon \right] \\
  (p_x+ip_y)(\beta_p-m) 
   \left[ \beta_p({\rm \Delta}\epsilon -2U_A -2\epsilon_0)
   +m\delta\epsilon \right] \\
   \delta\epsilon
   [(\epsilon_- +\beta_p-U_A)p_z^2 -2\beta_p^2(\beta_p-m)]
   +2\beta_p(\beta_p-m)
   \left[ (\epsilon_- -U_A)(\epsilon_+-U_A) 
   -\epsilon_0^2 \right] \\
   -p_z(p_x+ip_y)(\beta_p-m) \delta\epsilon
  \end{array} \right). \nonumber
\end{equation}
%

\section{Decomposition of $B_{\lowercase{n}}({\bf p})$ 
               in terms of the Dirac gamma matrices}
The operator $B_n(\vp)$ in eq.~(\ref{Bn}) consists of 
some gamma matrices;
 it is a linear combination of 
$\mbox{\boldmath $1$}$, 
$\mbox{\boldmath $\gamma$}$, $\gamma_5\mbox{\boldmath $\gamma$}$, 
$\sigma_{01}$ and $\sigma_{02}$, in the diquark field  
$\bar{\psi_c}B_n\psi=\psi^T C B_n \psi$.
The last two matrices give the tensor diquark fields,  
while these terms have no influence on the gap equation (\ref{gap1i})  
due to axial symmetry of the Fermi seas around the $p_z$ axis:  
the integration of $B_n({\vp})$ with respect to the azimuthal angle $\phi_p$ 
in eq.~(\ref{gap1i}) gives
\begin{equation}
\tilde{B}_n({\vp})\equiv
\int^{2 \pi}_0 \frac{{\rm d}\phi_p}{2 \pi} 
B_n({\vp})=
\frac{p_t}{4 |\epsilon_n({\vp})| \beta_p}
\left[ (-1)^n p_z \gamma_3+ (-1)^n m \mbox{\boldmath $1$}
+\beta_p \gamma_5 \gamma_3 \right]. \label{deco1}
\end{equation} 
Thus tensor terms disappear 
because they are proportional to $\exp{(i\phi_p)}$ in $B_n(\vp)$. 

The first term in the right hand side has also no contribution after 
symmetric integration with respect to $p_z$.
The remainders, $\{{\bf 1}, \gamma_5 \gamma_3\}$,  
imply the pseudo-scalar ($J^P=0^-$) and vector ($J^P=1^-$)
diquark pairings in terms of the notation in ref.~\cite{BL}.
Please note that the CSC gap (\ref{deco1}) results in 
a linear combination of different angular momentum pairs 
$0^-$ and $1^-$ because of the lack of rotation symmetry.
 
Since the diquark fields $\psi^TC({\bf 1}, \gamma_5 \gamma_3) \psi$ 
contain the off-diagonal matrices 
which connect the lower component with the upper one of the Dirac spinors, 
these pairings vanish in the non-relativistic limit 
or in the limit $m\rightarrow \infty$. Hence $B_n({\bf p})$ 
resembles $P$- wave
pairing as is seen in eq.~(50), although it has no
correspondence in the non-relativistic limit: 
the gap function for (\ref{deco1}) has the nodes (vanishing at $\theta=0,\pi$) 
due to the factor $p_t$, 
which is similar to $^3P$-pairing in the liquid $^3$He - A phase, 
but these nodes are entirely attributed to the genuine relativistic effect.
This property survives even
in the limit, $U_A \rightarrow 0$. 

From eq.~(\ref{deco1}) 
we can also obtain the relation appearing in eq.~(\ref{MN}); 
\begin{equation}
\gamma_\mu\gamma_0\tilde{B}_n({\vp})\gamma_0\gamma^\mu=
2 \tilde{B}_n({\vp}) +2 m\left\{ m+(-1)^n \beta_p \right\}. 
\end{equation} 
%
\section{Parameterization of the gap function}
In this Appendix 
we derive the parameterization (\ref{GAP1}).
The gap equation (\ref{para}) is expanded as
\begin{eqnarray}
\Delta_\pm(k) &=& \frac{2}{3}\tilde{g}^2
\int \frac{{\rm d}^3p}{(2\pi)^3}
\frac{k_t}{2\epsilon_\pm (k)} \left[ 
   \frac{p_t}{\epsilon_+(p)}
   \left(\pm \frac{2m^2}{\beta_k\beta_p}+1\right) \frac{\Delta_+(p)}{2 E_+(p)}
  +\frac{p_t}{\epsilon_-(p)}
   \left(\mp \frac{2m^2}{\beta_k\beta_p}+1\right) \frac{\Delta_-(p)}{2 E_-(p)} \right]. 
\end{eqnarray}
Introducing $\hat{\Delta}_\pm (k)$ through the equation,
\begin{equation}
\Delta_\pm (k)=\frac{k_t}{\epsilon_\pm (k)}\hat{\Delta}_\pm (k),
\end{equation}
we obtain the ``gap'' equation for $\hat{\Delta}_\pm (k)$,
\begin{equation}
\hat{\Delta}_\pm (k) = \frac{2}{3}\tilde{g}^2
\int \frac{{\rm d}^3p}{(2\pi)^3}
\frac{p_t^2}{4} \left[ 
  \mp \frac{2m^2}{\beta_k \beta_p}
    \left(\frac{\hat{\Delta}_-(p)}{\epsilon_-(p)^2 E_-(p)} 
         -\frac{\hat{\Delta}_+(p)}{\epsilon_+(p)^2 E_+(p)} \right) 
         +\frac{\hat{\Delta}_-(p)}{\epsilon_-(p)^2 E_-(p)} 
         +\frac{\hat{\Delta}_+(p)}{\epsilon_+(p)^2 E_+(p)} \right]
\label{C1}.
\end{equation}
Then we find the following properties, 
\begin{equation}
\hat{\Delta}_-(k)+\hat{\Delta}_+(k)=F ~~~\mbox{and}~~~ 
\hat{\Delta}_-(k)-\hat{\Delta}_+(k)=R \times \frac{m}{\beta_k},
\end{equation}
where $F$ ($R$) is a constant which characterize the symmetric (asymmetric) 
combinations of the gap functions $\hat{\Delta}_{\pm}$. 
Thus we can further parameterize $\hat{\Delta}_s(k)$ as 
\begin{equation}
\hat{\Delta}_\pm (k)=\mp \frac{m}{\beta_k} R + F \label{C2}.
\end{equation}
Substituting the above formula into eq.~(\ref{C1}), 
one can obtain the coupled equations for $F$ and $R$, eqs.~(\ref{eqF})
and (\ref{eqR}).
\section{Fierz transformation}
We present the Fock exchange energy term by the OGE interaction by the
use of  
the Fierz transformation.
The Green function with vertices in the right-hand side of eq.~(\ref{self1}) 
can be expanded as 
\begin{eqnarray}
\sum_a \left(\Gamma_a iG_{11} (p) \Gamma_a \right)_{ij} 
&=&
\sum_a (\Gamma_a)_{ii'} 
\langle \psi(p)_{i'} \bar{\psi}(p)_{j'} \rangle 
(\Gamma_a)_{j'j}
= 
\sum_{ab} C_{ab} (\Gamma_b)_{ij} {\rm Tr}(G_{11}\Gamma_b) \\ 
\mbox{with}~~
\Gamma_a &\equiv& {\gamma_\mu \otimes {\bf 1}_{flavor} \otimes \lambda_{color}} 
~~ \mbox{and}~~
(\Gamma_a)_{ii'} (\Gamma_a)_{j'j} =  
\sum_b C_{ab} (\Gamma_b)_{ij} (\Gamma_b)_{j'i'} \label{Fierz1},
\end{eqnarray}
where $\{C_{ab}\}$ are coefficients of a Fierz transformation (\ref{Fierz1})
for the Dirac matrices, the 
identity matrix in the flavor space and the Gell-Mann matrices in the 
color space;
\begin{eqnarray}
(\gamma_\mu)_{ii'} (\gamma^\mu)_{j'j} &=&  
\delta_{ij}  \delta_{j'i'} 
-\frac{1}{2}(\gamma_\mu)_{ij} (\gamma^\mu)_{j'i'}
-\frac{1}{2}(\gamma_5\gamma_\mu)_{ij} (\gamma_5\gamma^\mu)_{j'i'}
+(i\gamma_5)_{ij} (i\gamma_5)_{j'i'}\label{notensor}\\
\delta_{ii'} \delta_{j'j} &=&
\frac{1}{2}\left[\frac{2}{N_f}\delta_{ij} \delta_{j'i'}  
+(\tau_a)_{ij} (\tau_a)_{j'i'}\right]\\
(\lambda_c)_{ii'} (\lambda_c)_{j'j} &=&
\frac{2}{N_c^2} (N_c^2-1)\delta_{ij} \delta_{j'i'}  
-\frac{1}{N_c} (\lambda_c)_{ij} (\lambda_c)_{j'i'}.
\end{eqnarray}
It is to be noted that there appears no tensor term in
eq.~(\ref{notensor}) due to chiral symmetry in QCD.
Thus, e.g., the coefficient for the color-singlet axial-vector 
self-energy 
reads $-4/9$ for $N_f=2$ and $N_c=3$. 

We also present a Fierz transformation for diquark fields.
The right hand side of eq.~(\ref{gap1}) can be expanded, 
in the similar way for $G_{11}$;  
\begin{eqnarray}
\sum_a (\bar{\Gamma}_a G_{21}(p) \Gamma_a)_{ij}
&=&
\sum_a (C \Gamma_a^T C^{-1})_{ii'}
\langle \psi_c(p)_{i'} \bar{\psi}(p)_{j'} \rangle
(\Gamma_a)_{j'j}
=
\sum_a (C)_{ik} (\Gamma_a)_{i'k} 
\langle \bar{\psi}(-p)_{i'} \bar{\psi}(p)_{j'} \rangle 
(\Gamma_a)_{j'j} \nonumber \\
&=& \sum_{ab} f_{ab} 
{\rm Tr}\left( G_{21}(p) C^{-1}\Gamma_b^T C^{-1} \right) (C\Gamma_b^TC^T)_{ij} \\
\mbox{with} && (\Gamma_a)_{i'k} (\Gamma_a)_{j'j} =
\sum_b f_{ab} (\Gamma_bC^*)_{i'j'} (C\Gamma_b)_{jk}, \label{Fierz2}
\end{eqnarray}
where $\{f_{ab}\}$ are coefficients of a Fierz transformation (\ref{Fierz2}) and 
are explicitly given as 
\begin{eqnarray}
(\gamma_\mu)_{i'k} (\gamma^\mu)_{j'j} &=&
(C^*)_{i'j'} (C)_{jk} 
-\frac{1}{2} (\gamma_\mu C^*)_{i'j'} (C\gamma^\mu)_{jk} 
-\frac{1}{2} (\gamma_\mu\gamma_5C^*)_{i'j'} (C\gamma^\mu\gamma_5)_{jk}
+(iC^*\gamma_5)_{i'j'} (iC\gamma_5)_{jk} \\
\delta_{i'k} \delta_{j'j} &=& \frac{1}{2} 
\left[ \frac{2}{N_f} \delta_{i'j'} \delta_{jk}  
+(\tau_a)_{i'j'} (\tau_a)_{jk}  \right]\\
(\lambda_c)_{i'k} (\lambda_c)_{j'j}&=&
\left( 1-\frac{1}{N_c} \right) 
\left[ \frac{2}{N_c} (\delta)_{i'j'} (\delta)_{jk} 
+ (\lambda_c^S)_{i'j'} (\lambda_c^S)_{jk} \right]
-\left( 1+\frac{1}{N_c} \right) 
(\lambda_c^A)_{i'j'} (\lambda_c^A)_{jk},
\end{eqnarray}
where $\{ \lambda_c^{S(A)} \}$ are 
symmetric (antisymmetric) matrices of $\{\lambda_c\}$.  
The present gap function, 
$\Delta({\bf p})=\sum_n B_n({\bf p}) \Delta_n({\bf p}) $ 
which is a linear combination of the gamma matrices, 
can be obtained by taking projection on $B_n({\bf p})$.



----------------------
\end{document}